\documentclass{aa}  

\usepackage{graphicx}
\usepackage{txfonts}
\usepackage[colorlinks,citecolor=blue]{hyperref}
\begin{document} 

\title{Enhanced particle acceleration in a pulsar wind interacting with a companion}

\titlerunning{Pulsar wind interacting with a companion}

\author{Valentina Richard-Romei\inst{1} \and Benoît Cerutti\inst{1}}

\institute{Univ. Grenoble Alpes, CNRS, IPAG, 38000 Grenoble, France\\
           \email{valentina.richard-romei@univ-grenoble-alpes.fr}\\
           \email{benoit.cerutti@univ-grenoble-alpes.fr}
           }

\date{Received 03 May 2024; accepted 16 June 2024}

 
\abstract
{Pulsar winds have been shown to be preferred sites of particle acceleration and high-energy radiation. Numerous studies have been conducted to better characterize the general structure of such relativistic plasmas in isolated systems. However, many pulsars are found in binary systems and there are currently no ab initio models available that would include both the pulsar magnetosphere and the wind of the pulsar in interaction with a spherical companion.}
{We investigate the interaction between a pulsar wind and a companion to probe the rearrangement of the pulsar wind, assess whether it leads to an enhancement of particle acceleration, and predict the high-energy radiative signature that stems from this interaction. We consider the regime where the companion is small enough to hold between two successive stripes of the wind.}
{We performed two-dimensional (2D) equatorial particle-in-cell simulations of an inclined pulsar surrounded by a spherical, unmagnetized, perfectly conducting companion settled in its wind. Different runs correspond to different distances and sizes of the companion.}
{We find that the presence of the companion significantly alters the structure of the wind. When the companion lies beyond the fast magnetosonic point, a shock is established and the perturbations are advected in a cone behind the companion. We observe an enhancement of particle acceleration due to forced reconnection as the current sheet reaches the companion surface.  Hence, high-energy synchrotron radiation is also amplified. The orbital light curves display two broad peaks reaching up to 14 times the high-energy pulsed flux emitted by an isolated pulsar magnetosphere. These effects increase with the growth of the companion size and with the decrease of the pulsar-companion separation.}
{The present study suggests that a pulsar wind interacting with a companion induces a significant enhancement of high-energy radiation that takes the form of an orbital-modulated hollow cone of emission, which should be detectable by galactic-plane surveys, possibly with long-period radio transient counterparts.}


\maketitle

\section{Introduction}

A few percent of galactic pulsars are found in binary systems \citep{Lorimer_2008,breton_2009}. While tight binary systems involving magnetically coupled neutron stars have been studied in the past (e.g., \citealt{2001MNRAS.322..695H, 2013PhRvL.111f1105P, 2019A&A...622A.161C, 2020ApJ...893L...6M}), fewer studies have drawn attention to the interaction between a pulsar and a companion lying in its wind. However, it is a generic problem that has many astrophysical applications starting from binary pulsar systems, such as: (i) pulsar-neutron star such as $\rm PSR  ~B1913+16$ \citep{Hulse_1975} or the double pulsar system $\rm PSR~J0737-3039$ \citep{Burgay_2003,Lyne_2004}; (ii) pulsar-black hole (no confirmed detection so far; however, see \citealt{Barr_2024} and \citealt{chen_2024}); (iii) pulsar-main sequence star, including spider pulsars such as $\rm PSR~B1957+20$ \citep{Fruchter_1988} or $\rm PSR~J2051-0827$ \citep{Stappers_1996}; and (iv) pulsar-white dwarf, such as $\rm PSR~J1141-6545$ \citep{Kaspi_2000}, $\rm PSR~J1909-3744$ \citep{Jacoby_2003}, $\rm PSR~J1738+0333$ \citep{Jacoby_2007}, and $\rm PSR~J0348+0432$ \citep{Antoniadis_2013}. Planets orbiting pulsars (see \citealt{Nitu_2022} and references therein) form another class of relevant applications from which we expect characteristic signatures, as suggested by \citet{mishra_2023},  analogously to the Solar system planets and moons (e.g., \citealt{Neubauer_1980} for the Jupiter-Io interaction). We should also consider the case of asteroids interacting with a pulsar wind, which have  been proposed to trigger repeating fast radio bursts \citep{Dai_2016,Mottez_2020,Decoene_2021}. As of today, the basics of such interactions remain uncertain and we are left with opened questions regarding the possible rearrangement of the magnetosphere, the strength and location of particle acceleration, and the high-energy emission originating from such systems. Ultimately, we consider whether a pulsar wind interacting with a companion may lead to a new class of long-period high-energy transients.

Modelling the interaction between a pulsar wind and a companion requires us to capture both the magnetosphere’s global dynamics and the microphysical processes of the relativistic plasma responsible for particle acceleration and non-thermal radiation. To do so, we resorted to global particle-in-cell (PIC) simulations. Many studies have been conducted in order to characterize the general structure of isolated pulsar magnetospheres, whether aligned \citep{2014ApJ...785L..33P, 2014ApJ...795L..22C, Cerutti_2015, 2015MNRAS.449.2759B, Hu_2022} or inclined \citep{2015ApJ...815L..19P, Cerutti_2016, Kalapotharakos_2018}. These models have played a major role in the understanding of the magnetospheric structure and the kinetic processes at play. They have demonstrated that a sizeable fraction of the pulsar spindown power can be dissipated into kinetic energy. Magnetic reconnection has been identified as the predominant cause of particle acceleration and localized in the wind current sheet beyond the pulsar light cylinder. Accelerated particles escaping the reconnection layers have been shown to emit synchrotron radiation, mainly from the inner parts of the wind, thereby reproducing pulsars’ magnetospheric radiative signatures \citep{Cerutti_2016, Philippov_2018, Kalapotharakos_2018}. To our knowledge, no simulations of a pulsar magnetosphere and a spherical companion lying in the pulsar wind have been realized, neither in the framework of magnetohydrodynamics nor in that of PIC. However, the regime where a millisecond pulsar interacts with a low-mass companion star (i.e., the spider pulsar regime) has been explored with 2D PIC simulations by focusing on the outer parts of the wind, where the pulsar wind was modeled by a plane parallel striped wind (see \citealt{Cortes_2022,cortes_2024}).

In this work, we aim to probe intermediate scales, namely, scales enabling us to simulate an inclined pulsar magnetosphere and its wind. We focus our study on a spherical unmagnetized companion settled in the pulsar wind, small enough so that its diameter is shorter than the wind stripes width, which excludes the regime of spider pulsars from our field of study. We investigate the structure of the magnetosphere, the enhancement of particle acceleration, and the high-energy emission arising from this interaction. Furthermore, we probe the impacts of the companion size and separation on these diagnostics.

In the following, we first introduce the numerical model employed in this study (Section~\ref{section:numerical_model}). Section~\ref{section:isolated_case} describes our reference case, the isolated pulsar magnetosphere, recalling the main characteristics of the striped wind and the main kinetic and observational features. Our results on the pulsar-companion interaction are reported in Section~\ref{section:results}, where we first describe the wind global dynamics before discussing particle acceleration and high-energy radiation. In Section~\ref{section:discussion}, we outline the main results and their implications, and we discuss future prospects.

\section{Numerical model} \label{section:numerical_model}
We resort to the ab-initio PIC method, which allows us to capture both the global pulsar magnetosphere as well as the kinetic processes at play. We employ the 2.5D version of the relativistic electromagnetic {\tt Zeltron} code \citep{Cerutti_2013,Cerutti_2019}, in spherical coordinates \citep{Cerutti_2015,Cerutti_2016}, restricted to the equatorial plane. We used nearly the same setup as the one presented in \citet{Cerutti_2017}, to which we added a companion in the pulsar wind. For the sake of completeness, the full setup is described below.

\subsection{Initial setup} \label{subsection:initial_setup}

\begin{figure}[bthp]
\centering 
\includegraphics[width=\columnwidth, keepaspectratio]{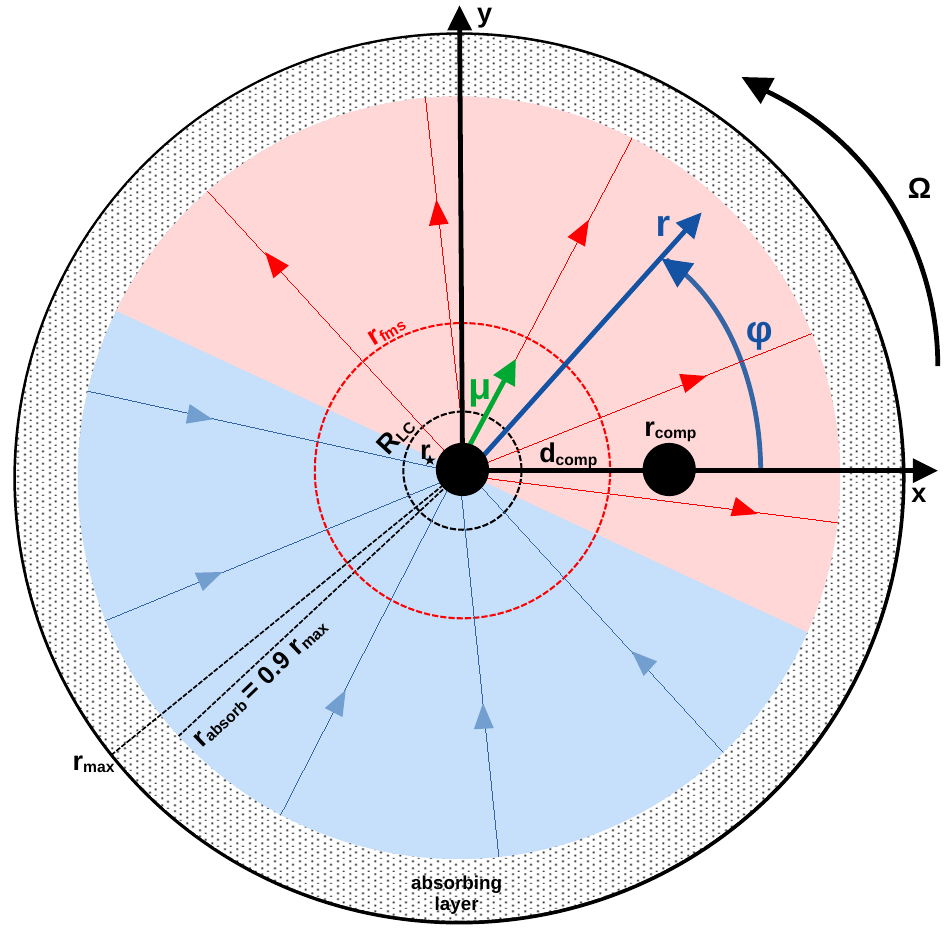}
\caption{Initial numerical setup. The pulsar, of a radius, $r_\star$ and angular velocity, $\Omega$, lies at the center of the spherical box of coordinates ($r$,$\phi$) and radius $r_{\rm max}$. The companion, of radius $r_{\rm comp}$, is settled at a fixed distance $d_{\rm comp}$ along the $\phi=0$ direction. Initially, the box is empty of particles. Magnetic field lines (blue and red arrows) are radial, directed outwards on one hemisphere and inwards on the other hemisphere, according to the split-monopole configuration. The plane inclination in between both hemispheres is perpendicular to the magnetic moment $\mathbf{\mu}$ shown in green. An absorbing layer is implemented at $r_{\rm absorb}=0.9 r_{\rm max}$. The light-cylinder radius $R_{\rm LC}$ is represented by the black dashed circle and the fast magnetosonic radius $r_{\rm fms}$ is given by the red dashed circle.}\label{fig:sketch_setup}
\end{figure}

The computational domain is a disk made of ($4096$ x $4096$) cells (see Fig.~\ref{fig:sketch_setup}). Particles are therefore confined in the $r\phi$-plane, but we keep the three spatial components of the electromagnetic field and of the particle velocities. The spherical grid is linear in $\phi$ and logarithmic in $r$, so as to keep the cell aspect ratio constant with radius. It is well suited for pulsar winds since the particles density and the field amplitudes are decreasing functions of the radius, meaning that the relativistic plasma skin depth, $c/\omega_p$ (where $\omega_p$ is the plasma frequency) and the Larmor radius, $r_L$, increase with radius. The surface of the star, $r=r_\star$, determines the inner boundary and the box extends radially until 24 $R_{\rm LC}$, where $R_{\rm LC}$ is the light-cylinder radius set at $R_{\rm LC}=3 r_\star$. The inner boundary absorbs all particles and an absorbing layer is implemented at $r_{\rm absorb}=0.9 r_{\rm max}$ for both particles and fields in order to mimic an open boundary after which no information can go back inwards \citep{Cerutti_2015,2015NewA...36...37B}. This assumption is reasonable given that the fast magnetosonic point lies well inside the outer boundary (see Section~\ref{subsection:companion_implementation}). 

The neutron star is modeled at the center of the box as a spherical perfect conductor with a constant angular velocity of $\mathbf{\Omega}_{\rm spin}=(c/R_{\rm LC})\,\mathbf{u}_{z}$, where the unity vector, $\mathbf{u_z}$, points in the out-of-plane direction. Initially, the star is in vacuum with magnetic field lines anchored at its surface. The implementation of a dipolar magnetic field for such equatorial configuration was proven to be inadequate (see \citealt{Cerutti_2017}). We therefore replace it by a split-monopole configuration \citep{1973ApJ...180L.133M,1999A&A...349.1017B}, for which the initial magnetic field is purely radial and reverses across the plane perpendicular to the magnetic moment $\mathbf{\mu}$, so as to ensure $\boldsymbol{\nabla} \cdot \mathbf{B}=0$ (see Fig.~\ref{fig:sketch_setup}). By construction of the equatorial setup, the magnetic axis must be inclined at an angle $\chi = \pi/2$. The perfect conductor condition applied to the pulsar surface in the corotating frame implies, by Lorentz transformation, a non-zero electric field in the simulation frame: 
\begin{eqnarray}
    \mathbf{E_\star}=-\frac{(\mathbf{\Omega}_{\rm spin}\times \mathbf{r_\star})\times \mathbf{B_\star}}{c} \,,
\end{eqnarray}
where $\mathbf{E_\star}$ and $\mathbf{B_\star}$ are the fields at the surface of the star.
This constraint starts the rotation of the magnetic field lines at $t=0$. 

The magnetosphere is initially empty and becomes progressively filled with electron-positron pairs that are evenly injected from the surface of the star at a rate of one macroparticle per cell per timestep for each species. They account for polar cap discharge and pair creation processes \citep{1971ApJ...164..529S, 1975ApJ...196...51R, 2013MNRAS.429...20T, 2020PhRvL.124x5101P}. Indeed, we neglected any other pair creation process away from the star surface,  given that pairs are mainly produced in the inner magnetosphere. The presence of ions (largely dominated by pairs in terms of number density) would not modify the magnetospheric structure and their radiative signature would be largely exceeded by pair losses. This justifies our choice not to model the ion extraction from the surface. We set a high surface multiplicity of $\kappa_\star=n_\star/n_{\rm GJ}=10$, where $n_\star$ is the density of the plasma injected at the surface of the star and $n_{\rm GJ}=\Omega B_\star/2\pi e c$ is the Goldreich-Julian density \citep{Goldreich_1969} at the surface of the star. This ensures that the plasma is sufficiently dense to screen the parallel electric field ($\mathbf{E}\cdot\mathbf{B}=0$). We also set a high plasma magnetization at the surface of the star: $\sigma_{\star}=B^2_{\star}/4\pi \Gamma_\star n_{\star} m_e c^{2}$, where $\Gamma_\star$ is the surface plasma bulk Lorentz factor. Having very high multiplicity and plasma magnetization (i.e.,~$\kappa\gg1$ and $\sigma\gg1$) ensures that the quasi force-free limit is achieved in the magnetosphere and in the wind (except deviations in the current sheet), meaning that the Lorentz force dominates over any other force: the conservation of momentum equation therefore reads $\rho \mathbf{E}+\mathbf{J}\times \mathbf{B}/c=0$, where $\rho$ and $\mathbf{J}$ are respectively the charge and current densities. Pair plasma is injected along the magnetic field lines with an initial velocity given by the drift velocity expected for the monopole solution in the force-free limit: $\mathbf{V}_D=c\mathbf{E}\times\mathbf{B}/B^{2}$.

The scale separation in the computing box is reduced by several orders of magnitude ($\sim 10^{2-4}$) compared to a realistic pulsar to resolve the plasma kinetic scales. Our simulations are therefore best suited for millisecond pulsars. The plasma skin depth $d_{\rm e}$ is resolved by $1.14$ cells ($\Delta r$) at the surface of the star (i.e., at its minimum value $d_{\rm e}^{\star}$), $(d_{\rm e}/\Delta r)_{\rm LC}\sim 10$ at $r=R_{\rm LC}$, and $d_{\rm e}$ globally increases with radius up to a resolution of $\sim 24$ cells at the outer boundary. The current sheet that forms in the magnetosphere (see Section~\ref{section:isolated_case}) has a width of the order of $d_{\rm e}$. At $r=R_{\rm LC}$, the Larmor radius resolution goes from $\sim 1$ cell in the wind to $\sim 70$ cells inside the current sheet. The Larmor radius resolution then gradually increases with radius up to $\sim 10$ cells in the wind at the edge of the box.

\subsection{Companion implementation}\label{subsection:companion_implementation}

We added a companion in the wind of the central pulsar (see Fig.~\ref{fig:sketch_setup}). For simplification purposes, the companion is chosen to be an unmagnetized perfect conductor, with no wind and no intrinsic spin. Particles hitting its surface are absorbed. We aim to study the impact of the binary separation and of the companion radius on the pulsar magnetosphere and wind. Therefore, we ran six simulations for different choices of separations ($d_{\rm comp}$) and radii ($r_{\rm comp}$). We refer to Table~\ref{tab:run_parameters} for the full set of parameters. The pulsar wind is globally made of two nested Archimedean-shaped stripes of alterning magnetic polarity, with a wavelength of $\lambda=2\pi R_{\rm LC}$, separated by current sheets (see Section~\ref{section:isolated_case}).
We restricted our study to companion sizes ensuring
\begin{eqnarray}
    \delta_{\rm cs} < r_{\rm comp} < \pi R_{\rm LC}\,,
\end{eqnarray}
where $\delta_{\rm cs}$ is the current sheet width. Indeed, we want the companion radius to be smaller than the semi-stripe wavelength so that the whole diameter holds between two successive reconnection layers. Such choice excludes the spider pulsar regime. Given our choice of companion separations and radii (see Table~\ref{tab:run_parameters}), the $\delta_{\rm cs}/r_{\rm comp}$ ratio varies from $0.15$ (run D2R1) to $0.9$ (run D9R05).

We probe different zones of the wind by changing the separation of the companion $d_{\rm comp}$. In particular, we wish to investigate the impact of the companion position with respect to the fast magnetosonic point, $r_{\rm fms}$. This point is defined as the radius for which the wind velocity exceeds the Alfvén speed \citep{Michel_1969,2009ASSL..357..421K}:
\begin{eqnarray}
    \Gamma_{\rm fms}=\left(\frac{B^{2}}{4\pi n m_e c^{2}}\right)^{1/3}=(\Gamma \sigma)^{1/3} \,,
\end{eqnarray}
where $\sigma$ is the plasma magnetization.
We need the fast magnetosonic point to be far enough from the light cylinder radius as well as from the outer boundary of the box. This constraint leads us to fix a magnetization of $\sigma_\star=250$ at the neutron star surface, meaning $\sigma_{\rm LC} \sim 60$ at the light-cylinder radius outside the current sheet.

The pulsar-companion separation is assumed to be constant in time. Indeed, even considering a compact object binary made of millisecond pulsars and for our closest separation (i.e., $d_{\rm comp}=2 R_{\rm LC}$), the merger time due to the emission of gravitational waves is $300$ times longer than the pulsars spin period ($P_{\rm spin}$) and $30$ times longer than their orbital period ($P_{\rm orb}$), according to the approximate analytical expression of merger time derived in \citet{2023MNRAS.524..426I}. The companion is settled at rest in the simulation. Even for the closest binary separation ($d_{\rm comp}=2 R_{\rm LC}$), the orbital period computed considering the Keplerian orbit of a millisecond pulsar binary is $10$ times longer than the pulsar spin period ($P_{\rm spin}$). This assumption, added to the perfect conductor condition, implies that $\mathbf{E}=\mathbf{0}$ at the surface and everywhere inside the companion.

\subsection{Fields and particle evolution}

Starting from the initial conditions given above, the PIC method allows us to couple the particles evolution along with time-dependent electromagnetic fields in a self-consistent way.  Firstly, the Boris push \citep{1991ppcs.book.....B} is used to evolve particles positions and velocities according to the Abraham-Lorentz-Dirac equation. This equation accounts for radiative energy losses by adding a radiation-reaction force ($\mathbf{f}_{\rm rad}$) to the Lorentz force:
\begin{equation}
    \frac{{\rm d}(\gamma m_e \mathbf{v})}{{\rm d} t}= q (\mathbf{E}+\boldsymbol{\beta} \times \mathbf{B})+ \mathbf{f}_{\rm rad} \,,
\end{equation}
where, for each particle, $\mathbf{v}=\boldsymbol{\beta}c$ is the 3-velocity, $q$ is the electric charge, and $\gamma$ is the Lorentz factor. The radiation-reaction force is implemented according to the Landau-Lifshitz formula \citep{1971ctf..book.....L}, in the \citet{2010NJPh...12l3005T} approximation (see \citealt{Cerutti_2016} for full details) as:
\begin{eqnarray}
    \mathbf{f}_{\rm rad}=\frac{2}{3}r^{2}_e [(\mathbf{E}+\boldsymbol{\beta}\times \mathbf{B})\times \mathbf{B}+(\boldsymbol{\beta}\cdot\mathbf{E})\mathbf{E}] \nonumber\\
    -\frac{2}{3}r^{2}_e \gamma^{2} [(\mathbf{E}+\boldsymbol{\beta}\times\mathbf{B})^{2}-(\boldsymbol{\beta}\cdot\mathbf{E})^{2}] \,\boldsymbol{\beta},
\end{eqnarray}
where $r_e=e^{2}/m_e c^{2}$ is the classical electron radius. Due to numerical costs, each simulated particle represents a large number of real particles, given by the weight, $w_k$, with the same $q/m$ ratio and therefore following the same trajectory in phase space. 
Secondly, charges and currents are deposited on the grid, based on an area-weighting deposition scheme. 
Thirdly, electromagnetic fields are evolved on the grid by solving Maxwell-Faraday and Maxwell-Amp\`ere equations through the finite-difference-time-domain algorithm \citep{1966ITAP...14..302Y}. While $\boldsymbol{\nabla} \cdot \mathbf{B} =0$ is automatically verified under these conditions, the conservation of charge is not ensured, due to machine truncation errors. We then solve the Poisson equation with the iterative Gauss-Seidel method every $25$ timesteps ($\Delta t$). The timestep is given by half the Courant-Friedrich-Lewy critical condition. We note that general relativistic corrections on the electrodynamics of the system are neglected in this work (see however \citealt{2015ApJ...815L..19P,Philippov_2018,2020ApJ...889...69C,Torres_2024}).
The full list of numerical and physical parameters chosen for the simulations is given in Table~\ref{tab:initial_params}.

 \begin{table}[ht!]
    \caption{Simulation parameters. Values evaluated at $R_{\rm LC}$ or $r_{\rm fms}$ are computed for the isolated pulsar magnetosphere.}
    \centering
    \begin{tabular}{lc}
        \hline
        \hline
        \noalign{\smallskip}
        Parameter & \quad Value \\
        \noalign{\smallskip}
        \hline
        \noalign{\smallskip}
        Number of cells & \qquad $4096\,(r) \times 4096\,(\phi)$ \\
        Inner boundary   & \qquad  $r_\star$ \\
        $R_{\rm LC}$  & \qquad 3 $r_\star$\\
        $r_{\rm absorb}$  & \qquad 24 $R_{\rm LC}$ \\
        $(d_e/\Delta r)_{\rm LC}$ & \quad 16.2   \\
        $\sigma_{\rm LC}$ & \qquad $60$  \\
        $P_{\rm spin} / \Delta t$ & \qquad  $4.3 \times 10^{4}$ \\
        $r_{\rm fms}$ & \qquad 5.1 $R_{\rm LC}$\\
        $\Gamma_{\rm fms}$ & \qquad 3.9\\
        $d^{\star}_{\rm e}/r_\star$ & \qquad  $1.8 \times 10^{-3}$\\ 
        Plasma composition & \qquad electrons and positrons \\
        Injection model & \qquad from the star surface \\
        \noalign{\smallskip}
        \hline
    \end{tabular}
    \label{tab:initial_params}
\end{table}

\subsection{Radiation modeling}\label{subsection:radiation_modelling}

Synchrotron radiation is the main source of high-energy emission beyond the light cylinder \citep{Cerutti_2016}. Each macroparticle of the simulation emits a macrophoton, representing a set of physical photons and emitting the following power \citep{1970RvMP...42..237B}:
\begin{eqnarray}
    \mathcal{P}_{\rm rad}=\frac{2}{3} r_e^{2} c \gamma^{2}\tilde{B}_\perp^{2}\,,
\end{eqnarray}
with power spectrum
\begin{eqnarray}
    \frac{{\rm d}\mathcal{P}_{\rm rad}}{{\rm d} \nu} = \frac{\sqrt{3}e^{3}{\Tilde{B}_\perp}}{m_e c^{2}} \left(\frac{\nu}{\nu_c} \right)\int^{+\infty}_{\nu/\nu_c} K_{5/3} (x) {\rm d}x \,,
\end{eqnarray}
where $K_{5/3}$ is the modified Bessel function of order $5/3$, $\nu$ is the radiation frequency, $\nu_c = 3e\Tilde{B}_\perp\gamma^{2}/4\pi m_e c$ is the critical frequency, and $\Tilde{B}_\perp$ is defined in Eq.~(\ref{eq:Btildeperp}). Around pulsars, the electric and magnetic field strengths are comparable. Instead of referring to the classical expression $B_\perp=B \sin\alpha$ (where $\alpha$ is the angle between $\mathbf{B}$ and the direction of the particle), valid for synchrotron radiation in a pure magnetic field, we use the effective perpendicular magnetic field strength, $\Tilde{B}_\perp$, that takes into account an arbitrary electromagnetic field (see \citealt{Cerutti_2016} for further details):
\begin{eqnarray}\label{eq:Btildeperp}
    \Tilde{B}_\perp = \sqrt{(\mathbf{E}+\boldsymbol{\beta}\times\mathbf{B})^{2} - (\boldsymbol{\beta} \cdot \mathbf{E})^{2}} \,.
\end{eqnarray}
Quantum electrodynamical effects reached above the critical magnetic field, $B_{\rm QED}=m_e^{2}c^{3}/\hbar e$, are neglected since we globally have $\gamma \Tilde{B}_\perp\ll B_{\rm QED}$ in the magnetosphere. Photons are emitted along the particles momentum. This is a good approximation in the presence of strong relativistic beaming since the emission cone has a semi-aperture angle of $\sim 1/\gamma \ll 1$. Photons then propagate freely at the speed of light, without interacting between themselves nor with the magnetic field. However, they are absorbed if they hit the star or the companion. General relativistic effects on the photon trajectories are neglected. We collect the photons on a screen at infinity.

In this problem, we focus on the orbital modulation of the light curves. As previously mentioned, the companion is at rest in the simulation, under the assumption that $P_{\rm orb} \gg P_{\rm spin}$. However, assuming a circular orbit, we can model orbital-modulated light curves by placing observers all around the box. An isolated pulsar emits a pulsed radiation over its spin period. To compute such pulsed light curves, we need to take into account the time delay between photons emitted at different locations of the box when they reach the screen \citep{Cerutti_2016}. Nevertheless, when we focus on orbital-modulated light curves, the delay times are way shorter than the orbital timescales and we do not need to consider them.

\section{Reference case:  Isolated pulsar magnetosphere} \label{section:isolated_case}

\begin{figure*}[tb]
\centering 
\includegraphics[width=0.98 \textwidth, keepaspectratio]{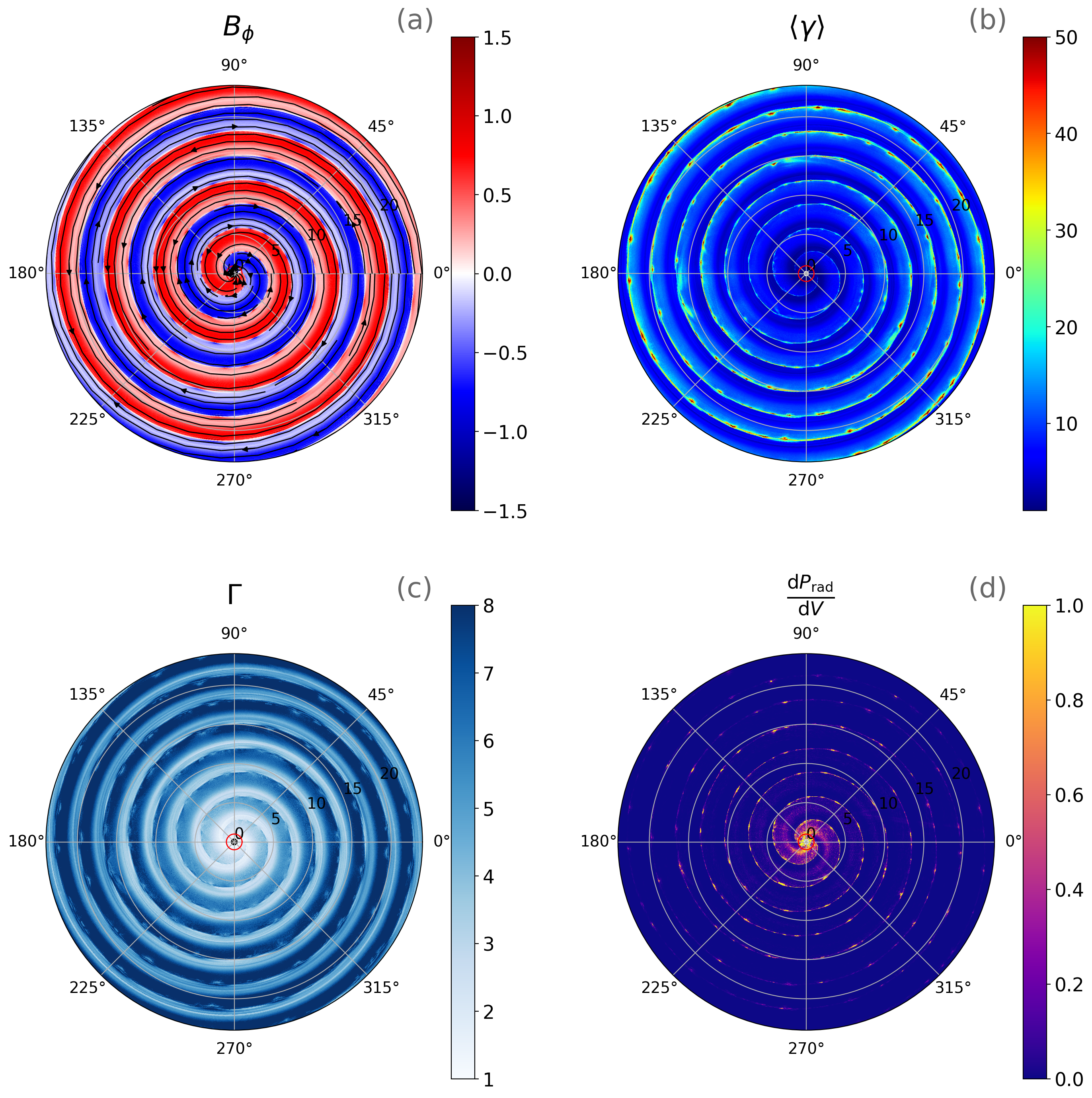}
\caption{Snapshot of the isolated pulsar simulation after $12.4$ pulsar spin periods. The central pulsar is represented by a black disk on left panels and a white disk on right panels, for better visualization purposes. \textit{From top-left to bottom-right:} Toroidal magnetic field $(B_\phi/B_\star)(R_{\rm LC} r/r_\star^{2})$, where solid lines are the magnetic field lines; the mean Lorentz factor of particles $\langle\gamma\rangle$; the bulk Lorentz factor of the plasma $\Gamma$; and the isotropic synchrotron power emitted by the particles per unit of volume $({\rm d}P_{\rm rad}/{\rm d}V)(r^{2}/r_\star^{2})$. Radii are given in units of the light cylinder radius, represented by the red circle.}\label{fig:ref_case}
\end{figure*}

\begin{figure*}[tb]
\centering 
\includegraphics[width=0.98 \textwidth, keepaspectratio]{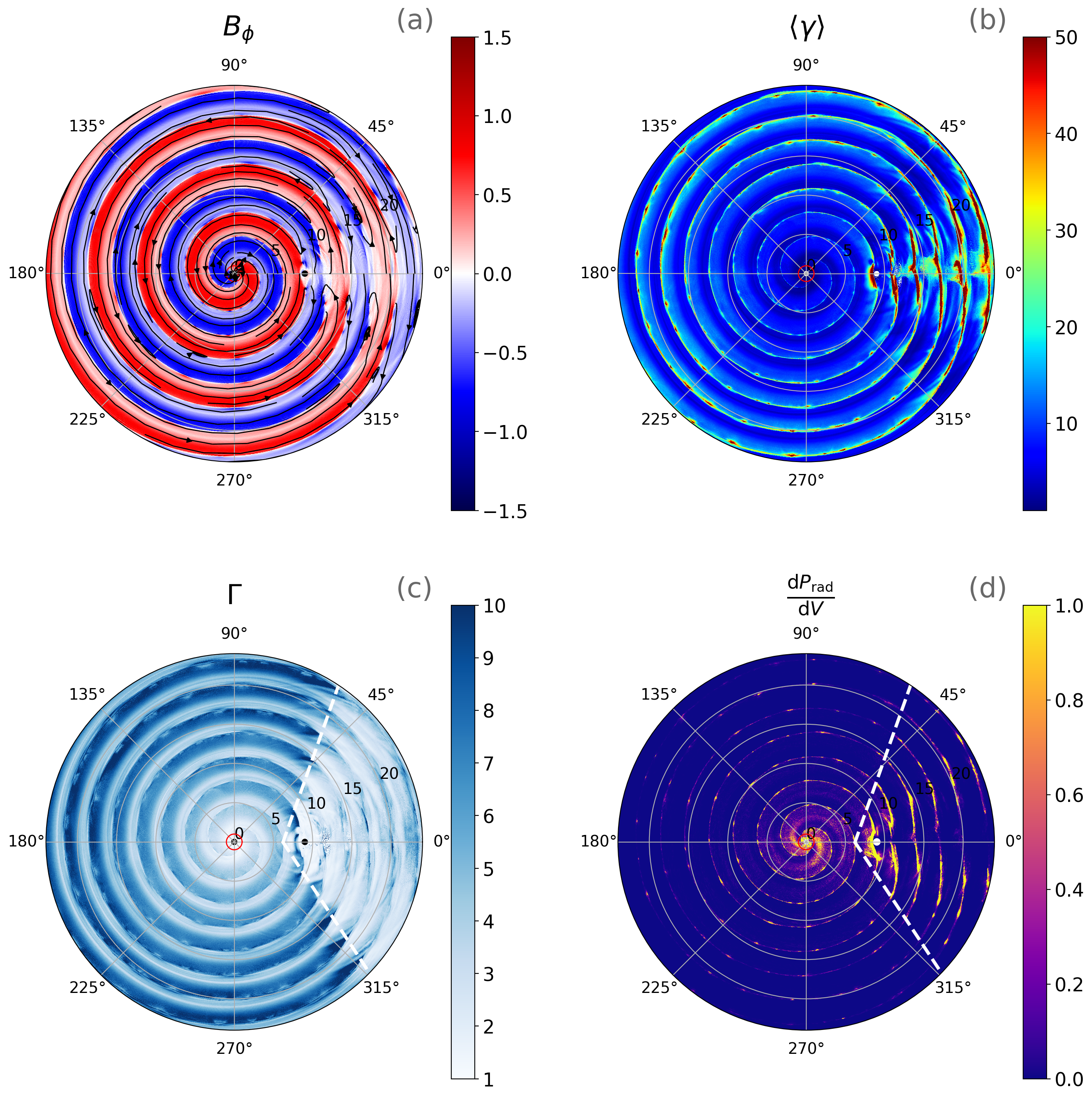}
\caption{Snapshot of the D9R1 run after $11.7$ pulsar spin periods. The central pulsar and the companion (of distance $d_{\rm comp}=9 R_{\rm LC}$ and size $r_{\rm comp}=r_\star$) are represented by black disks on left panels and white disks on right panels, for better visualization purposes. \textit{From top left to bottom right:}  Toroidal magnetic field $(B_\phi/B_\star)(R_{\rm LC} r/r_\star^{2})$, where solid lines are the magnetic field lines; the mean Lorentz factor of particles, $\langle\gamma\rangle$; the bulk velocity of the wind, $\Gamma$; the isotropic synchrotron power emitted by the particles per volume unit $({\rm d}P_{\rm rad}/{\rm d}V)(r^{2}/r_\star^{2})$. Radii are given in units of the light cylinder radius, represented by the red circle. The white dashed lines indicate the aperture angle of the shocked zone appearing on panel (c); we report them in panel (d) for comparison with the emitting hollow cone.}
\label{fig:companion_case}
\end{figure*}

The initial split-monopole condition represented in Figure~\ref{fig:sketch_setup} quickly evolves as the magnetic field lines start rotating at $t>0$ to ensure the perfect conductor condition of the star (see Section~\ref{subsection:initial_setup}). Pairs injected from the surface gradually fill the box. The magnetosphere converges to a stationary solution after $\sim~4.5 P_{\rm spin}$ (Fig.~\ref{fig:ref_case}).  Within the light cylinder, the magnetic equatorial regions are characterized by closed field lines that co-rotate with the star and trap the plasma, whereas open field lines of opposite polarities originating from the polar caps escape the light cylinder due to the rotation of the star and reconnect along the current sheet \citep{2009ASSL..357..421K,2012SSRv..173..341A,P_tri_2016,Cerutti_revue}.
When the magnetic axis is inclined with respect to the rotation axis (recall $\chi=\pi/2$ here), the current sheet takes the shape of an oscillatory structure in $\theta$, with wavelength $2\pi R_{\rm LC}$ and angular aperture $2 \chi$, referred to as the ``striped wind'' \citep{1990ApJ...349..538C,1999A&A...349.1017B}. A 2D equatorial cut of the current sheet results in two Archimedean spirals separated by stripes $\pi R_{\rm LC}$ wide of alterning magnetic field polarities, conveying the outflowing cold relativistic wind (see Fig.~\ref{fig:ref_case}a). 

Figure~\ref{fig:ref_case}c shows a snapshot of the bulk Lorentz factor of the wind at $t=12.4 P_{\rm spin}$. On average, the bulk Lorentz factor of the wind increases almost linearly until it reaches $\Gamma = 3.9$ at the fast magnetosonic point, $r_{\rm FMS} = 5.1 R_{\rm LC}$. After this point, the wind continues to accelerate at a slower rate up to $\Gamma \sim 7.5$ at the box edge. Due to magnetic reconnection locally attracting particles, the wind is slowed down on the leading edge of the current sheet whereas its acceleration increases on the trailing edge of the spiral.
Right after its formation near the light cylinder, the current sheet fragments due to the relativistic tearing instability \citep{1979SvA....23..460Z}, giving rise to a chain of plasma overdensities confined in magnetic islands, called plasmoids \citep{Cerutti_2017,Cerutti_2020,2023ApJ...943..105H}. Plasmoids then gradually grow by merging with each other while flowing outwards along the spirals. In between plasmoids, short current layers (referred to as X-points) allow for relativistic magnetic reconnection \citep{2010PhRvL.105w5002U}, which is the main physical process responsible for magnetic energy dissipation into particles kinetic energy \citep{2014ApJ...785L..33P,2014ApJ...795L..22C,2015ApJ...801L..19P,Cerutti_2015,Cerutti_2017,Cerutti_2020,2015MNRAS.449.2759B,Hu_2022}. About $24 \%$ of the Poynting flux reservoir (i.e., the spindown power extracted from the star) is consumed via magnetic reconnection and converted into kinetic energy between the light cylinder and the outer part of the box. Reconnection is particularly efficient at accelerating particles for $r < 2 R_{\rm LC}$. Accelerated particles escaping the X-points are trapped by plasmoids. Figure~\ref{fig:ref_case}b shows the mean particle acceleration ($\langle\gamma\rangle$).

A significant fraction of kinetic energy is radiated away through synchrotron emission (see Fig.~\ref{fig:ref_case}d). As expected, non-thermal radiation is emitted from the current sheet and decays with radius given that the density and $\Tilde{B}^{2}_\perp$ decrease with radius as $1/r^{2}$. In the end, $0.7 \%$ of the spindown power is converted into high-energy radiation. Due to time delay effects, the synchrotron radiation is pulsed, with two short bright pulses per stellar spin period \citep{1996A&A...311..172L,10.1111/j.1365-2966.2012.21238.x,Arka_2013,Cerutti_2016,Philippov_2018,Kalapotharakos_2018,2023ApJ...954..204K}. We defined a ``reference flux,'' computed by averaging this pulsed radiation over one pulsar spin period, to later compare it with the orbital-modulated light curves computed in the presence of the companion (see Section~\ref{subsection:em_signature}).

\section{Pulsar-companion interaction}\label{section:results}

\begin{figure*}[tb]
\centering 
\includegraphics[width=0.98 \textwidth, keepaspectratio]{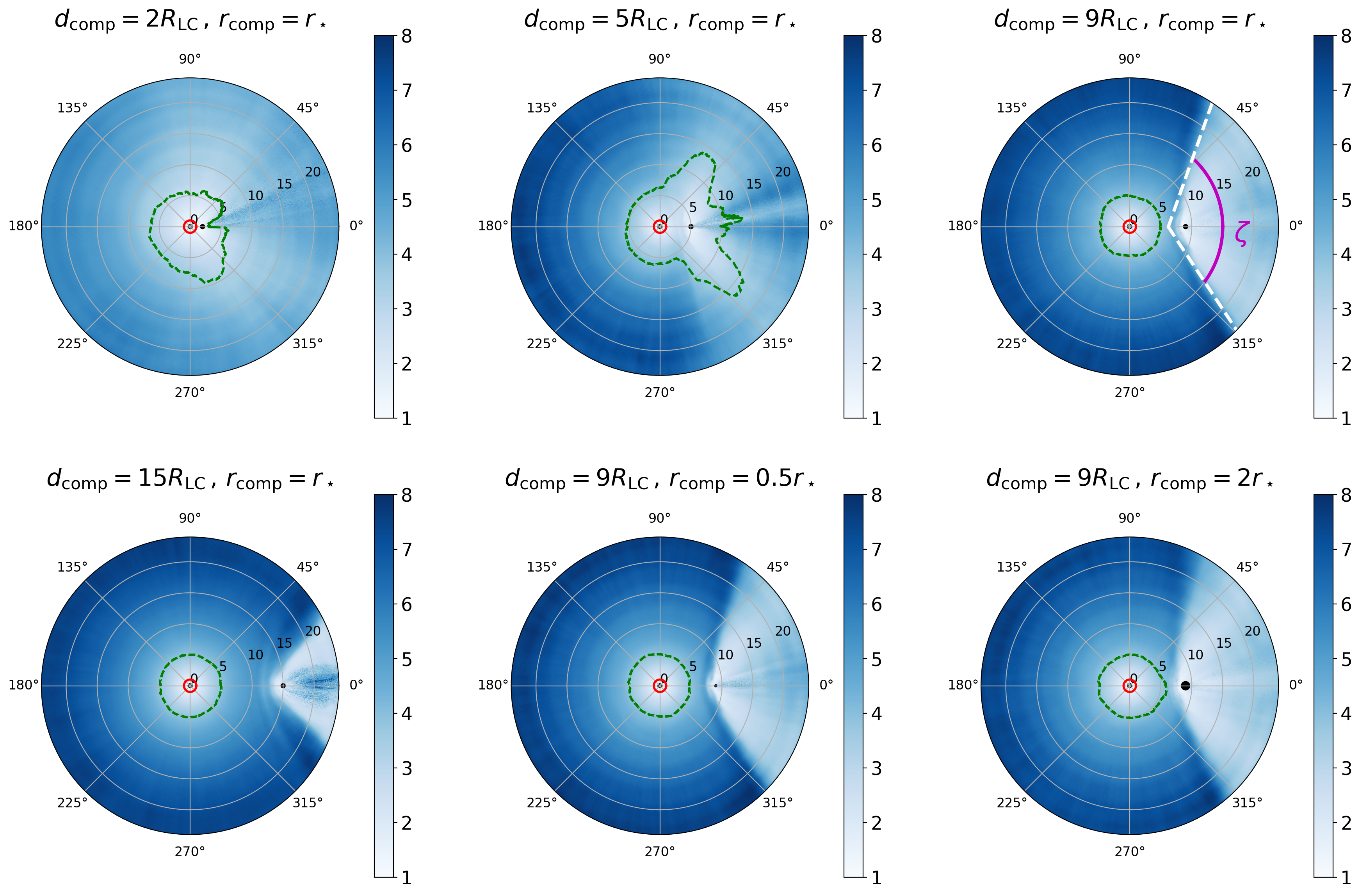}
\caption{Bulk Lorentz factor of the wind ($\Gamma_{\rm wind}$) averaged over several spin periods, for each run. Radii are given in units of the light-cylinder radius, represented by the red circle. The fast magnetosonic surface is represented by the dashed green curve. When $d_{\rm comp} > r_{\rm fms}$, a shock  of angular aperture $\zeta$ (see upper right panel) is established. Otherwise, no shock appears and the whole magnetosphere is slowed down.}\label{fig:gamma_bulk}
\end{figure*}

We carry out a parametric study in order to probe the impact of the companion on the pulsar wind, depending on the binary separation ($d_{\rm comp}$) and on the companion size ($r_{\rm comp}$). Table \ref{tab:run_parameters} details the parameters chosen for the six simulations we ran.

 \begin{table}
    \caption{Parameters chosen for each run.}
    \centering
    \begin{tabular}{lcr}
        \hline
        \hline
        \noalign{\smallskip}
        Run & \qquad $d_{\rm comp}\, (R_{\rm LC})$ & \qquad $r_{\rm comp}\, (r_{\star})$ \\
        \noalign{\smallskip}
        \hline
        \noalign{\smallskip}
        D2R1          & \qquad $2$ & \qquad $1$ \\
        D5R1      & \qquad $5$ & \qquad $1$ \\
        D9R05      & \qquad $9$ & \qquad $0.5$ \\
        D9R1        & \qquad $9$ & \qquad $1$\\
        D9R2 & \qquad $9$ & \qquad $2$ \\
        D15R1      & \qquad $15$ & \qquad $1$ \\
        \noalign{\smallskip}
        \hline
    \end{tabular}
    \label{tab:run_parameters}
\end{table}

\subsection{Wind dynamics}

Figure \ref{fig:companion_case} presents the same diagnostics as the ones we show for the isolated pulsar (Section~\ref{section:isolated_case}), but in the presence of a companion at $d_{\rm comp}= 9 R_{\rm LC}$ and $r_{\rm comp}=r_{\star}$ (run D9R1).
The conductor is settled beyond the fast magnetosonic point ($r_{\rm fms}\sim 5 R_{\rm LC}$) for this run. We observe an alteration of the magnetosphere (Fig.~\ref{fig:companion_case}, panel a), but (as expected) the perturbations are advected by the pulsar wind and, therefore, they stay in a cone behind the companion. On the contrary, when $d_{\rm comp} \leq r_{\rm fms}$ (runs D2R1 and D5R1), perturbations propagate faster than the outflowing wind and the whole magnetosphere is affected. As the wind reaches the companion, its bulk Lorentz factor sharply drops along a wider cone, indicating the presence of a shock (see Fig.~\ref{fig:companion_case}c). Given the strong magnetization of the magnetosphere, we did not expect large discontinuities in the magnetic field, nor in the density, that would provide evidence of a significant compression \citep{1984ApJ...283..694K}.

Figure \ref{fig:gamma_bulk} presents, for each of our simulations, the bulk Lorentz factor of the wind averaged over several spin periods of the pulsar, to which we added the fast magnetosonic surfaces. Averaging over several spin periods enables us to get rid of the wind striped structure and of local inhomogeneities, so as to keep only the overall bulk motion. We see that for pulsar-companion separations larger than the radius of the fast magnetosonic surface (lying at $r_{\rm fms} \sim 5 R_{\rm LC}$ according to the pulsar parameters set in this study), a shock is systematically established. However, when the companion is settled before the fast magnetosonic surface $d_{\rm comp} \leq 5 r_{\rm fms}$ (i.e., for runs D2R1 and D5R1), no shock appears. The wind is slowed down over the whole box, more isotropically and more intensely as the orbital separation decreases. We notice that, in this case, the fast magnetosonic surface gets disrupted but keeps encompassing the companion. The local enhancement of $\Gamma$ that appears in the wake of the companion is due to artefacts of the very low plasma density.

\begin{figure*}[tb]
\centering 
\includegraphics[width=0.16 \textwidth, keepaspectratio]{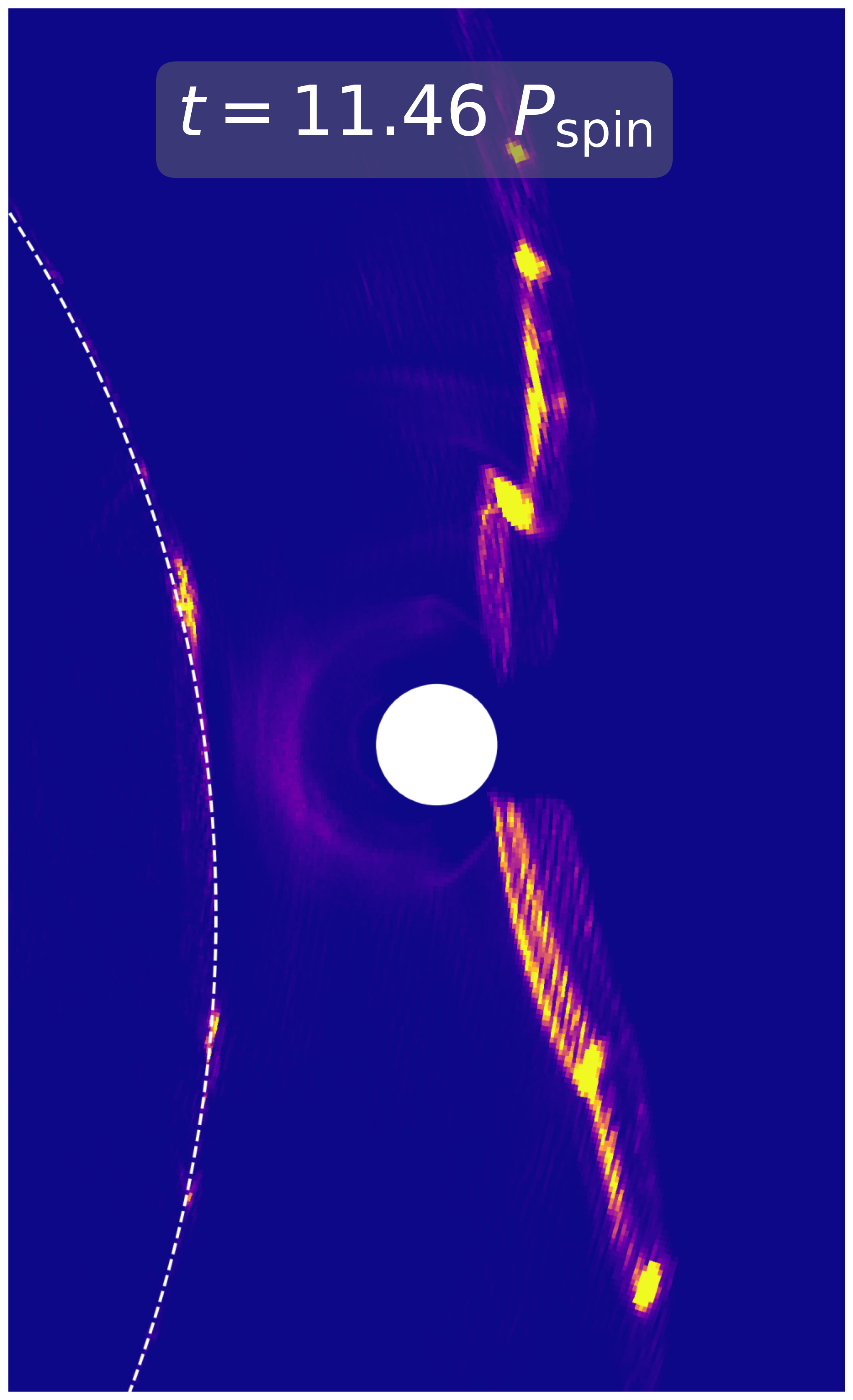}
\includegraphics[width=0.16 \textwidth, keepaspectratio]{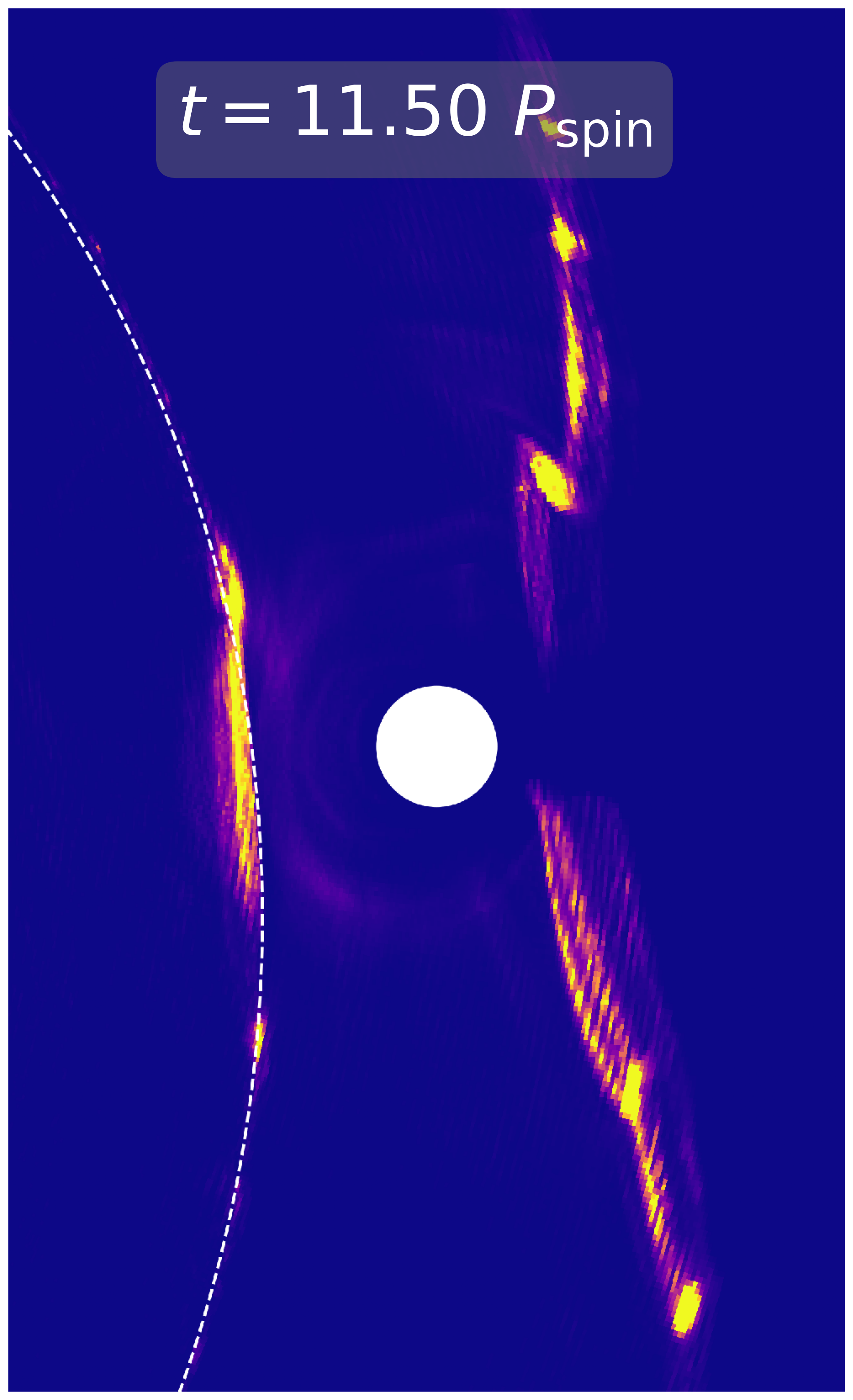}
\includegraphics[width=0.16 \textwidth, keepaspectratio]{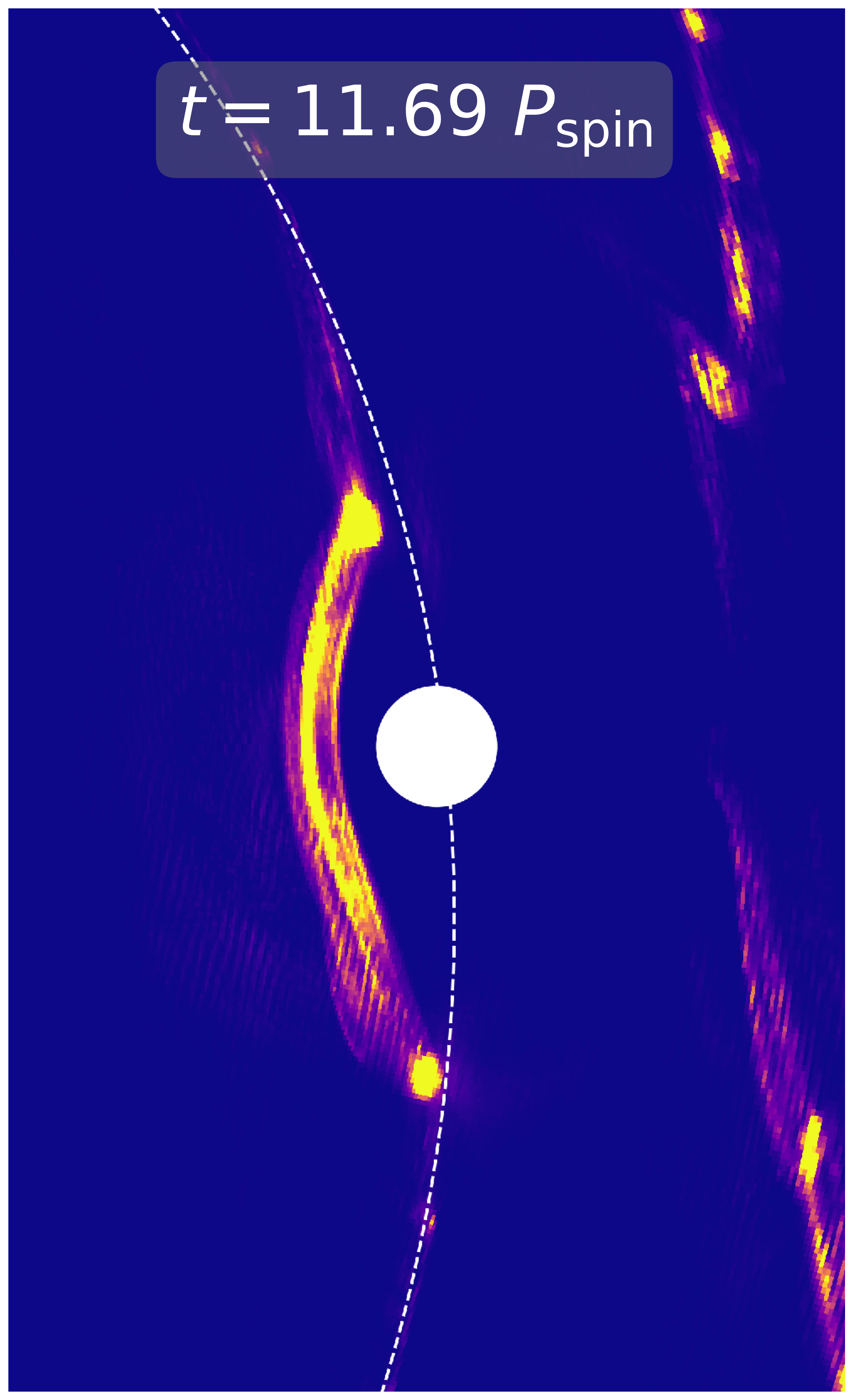}
\includegraphics[width=0.16 \textwidth, keepaspectratio]{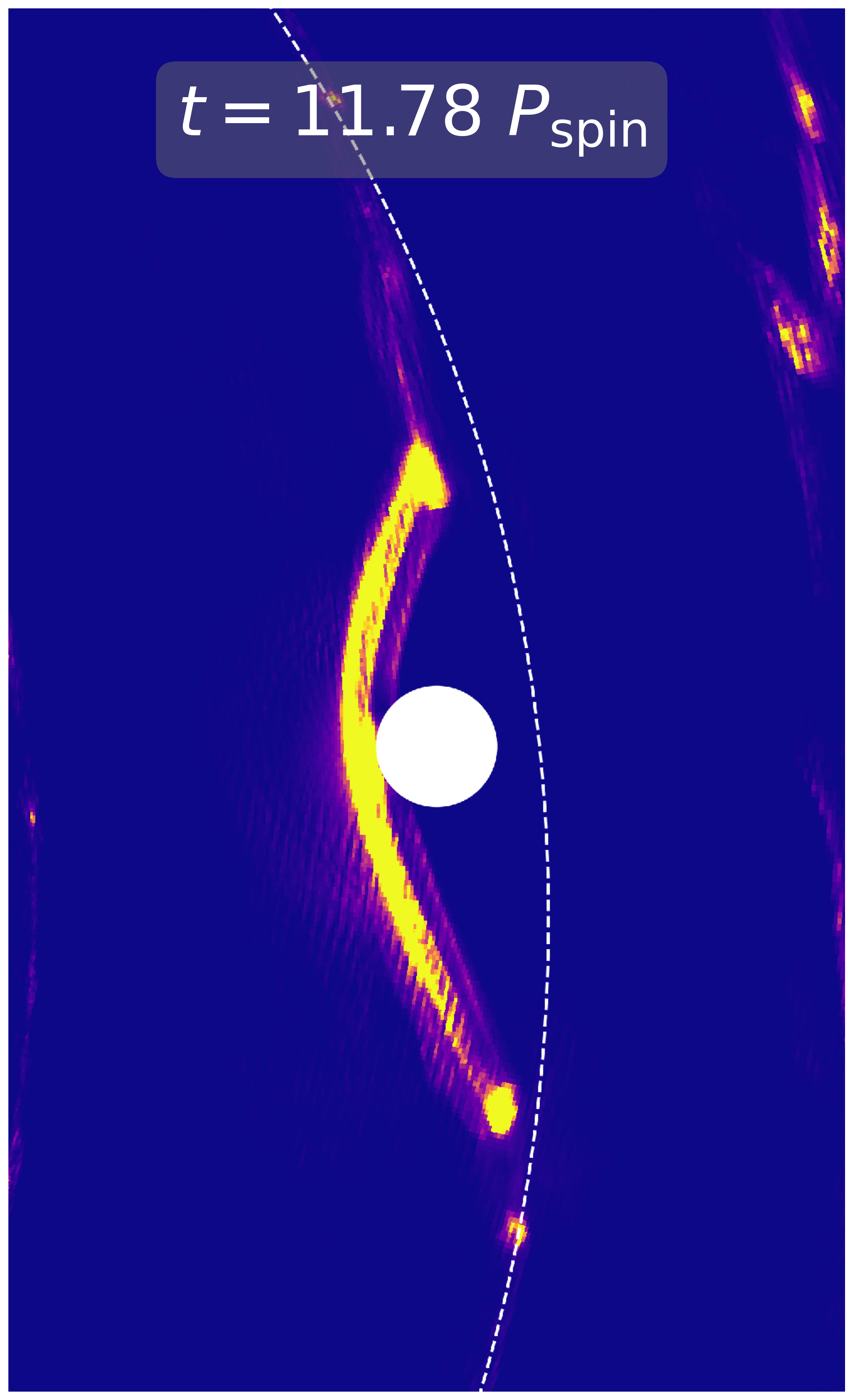}
\includegraphics[width=0.16 \textwidth, keepaspectratio]{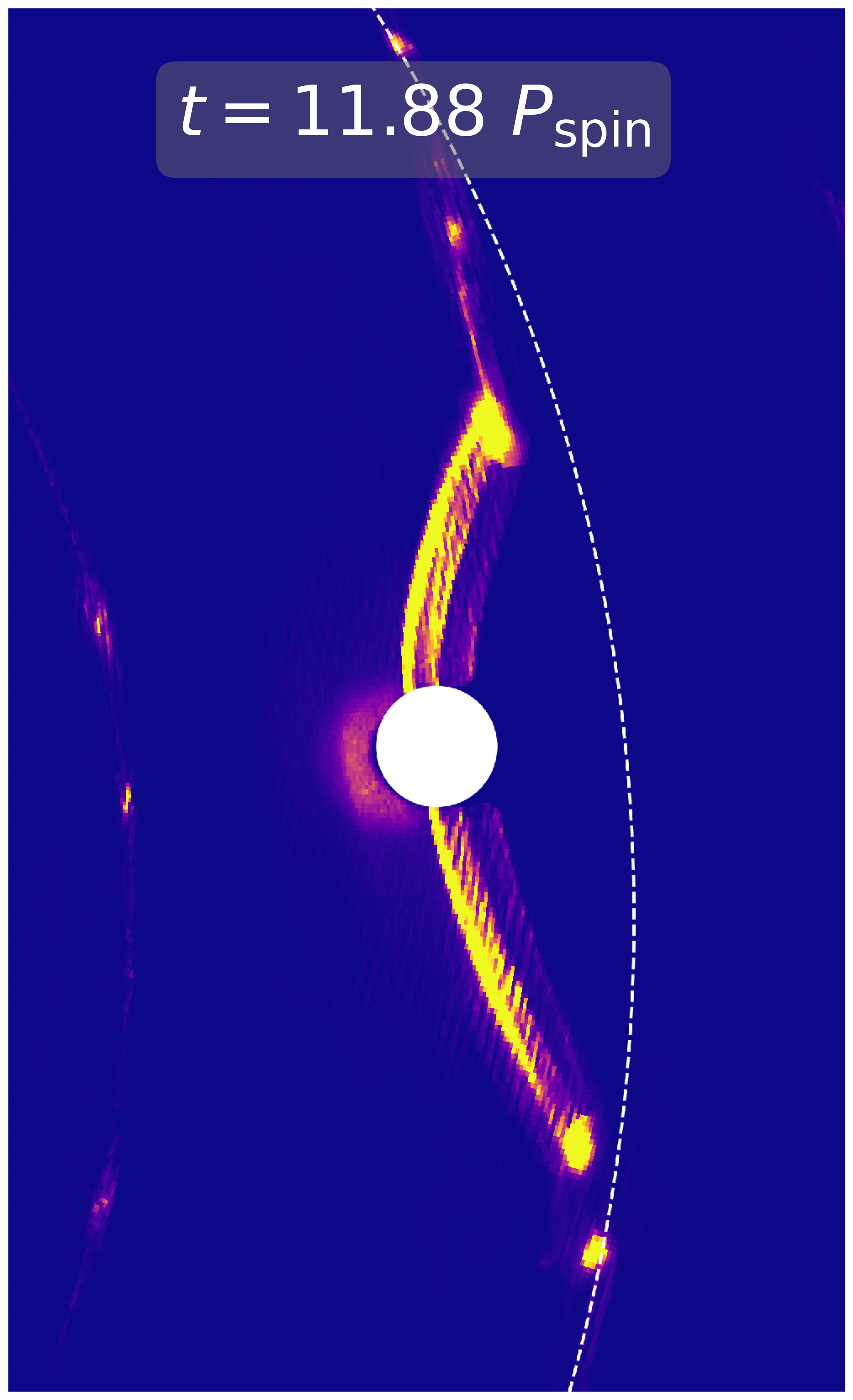}
\includegraphics[width=0.16 \textwidth, keepaspectratio]{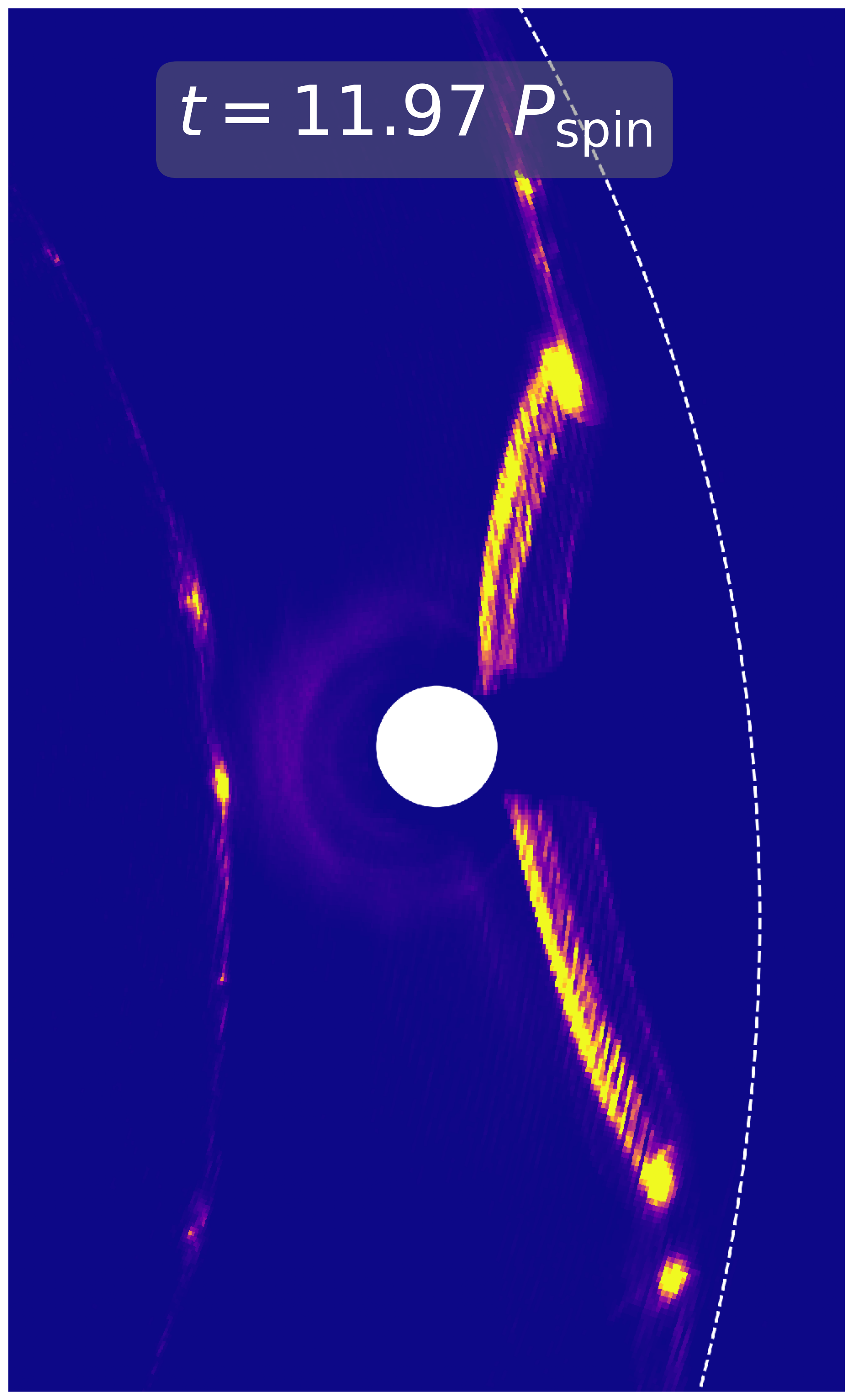}
\caption{Synchrotron power per unit of volume (${\rm d}P_{\rm rad}/{\rm d}V$) on successive snapshots of run D9R1, zoomed-in around the companion, during a current sheet-companion interaction. The white dashed line is an Archimedean spiral tracing the geometry that the current sheet (the one we focus on) would adopt in the absence of the companion. The current sheet flows almost radially at $d_{\rm comp}=9 R_{\rm LC}$. It bends and lights up when reaching the companion surface and finally breaks apart.}\label{fig:current_sheet}\label{fig:current_sheet}
\end{figure*}

When a shock is formed in a plane-parallel and uniform flow, the cone aperture angle ($\zeta$) can be related to the relativistic Alfvénic Mach number ($\mathcal{M}$) computed at the apex of the cone via the following relation:
\begin{eqnarray}\label{eq:mach}
    \sin \left(\frac{\zeta}{2}\right) \sim \frac{1}{\mathcal{M}}
,\end{eqnarray}
where $\mathcal{M}=\beta\Gamma/\beta_{\rm A}\Gamma_{\rm A}$, given that $\beta_{\rm A}=\sqrt{\sigma/(1+\sigma)}$ \citep{2009ASSL..357..421K} and $\Gamma_{\rm A}$ are  the Alfvén velocity and the Alfvén Lorentz factor, respectively. The measured values from the simulation (see $\zeta$ on the top right panel of Fig.~\ref{fig:gamma_bulk}) are in reasonably good agreement with the law presented in Eq.~(\ref{eq:mach}), with relative errors of less than $28 \%$.  The concordance is particularly remarkable given the high inhomogeneity of the striped wind and the tilt in the cone direction, which lead to different wind velocities reached for a same distance from the apex of the cone. Increasing the companion radius ($r_{\rm comp}$) leads to an increase of the cone aperture angle ($\zeta$). The altered part of the magnetosphere (Fig.~\ref{fig:companion_case} panel a) displaying enhanced particle acceleration and radiation lies inside the shocked zone, so that the aperture angle of the shocked cone also gives an upper boundary to the aperture of the emitting cone (see Fig.~\ref{fig:companion_case}d).

\subsection{Particle kinetics}

The presence of the companion in the magnetosphere of the pulsar enhances particle acceleration regardless of the binary separation. This comes from the compression of the magnetic field lines reaching the companion, which results in a forced reconnection. Indeed, as can be seen in Figure \ref{fig:current_sheet}, the forward part of the current sheet is first slowed down while approaching the companion (starting from a distance of about $1 r_{\rm comp}$). It then bends backwards but ends up reaching the companion surface as the current sheet continues to progress radially.  After the companion surface is reached, the current sheet eventually breaks apart around the obstacle and the two branches of enhanced particle acceleration flow outwards, creating two radial lines of enhanced particle acceleration. In fact, the whole cone between these two lines represents a favorable zone for particle acceleration (see Fig. \ref{fig:companion_case} panel b). However, the density is extremely low in the wake of the companion, so that most of the highly accelerated particles flow along the borders of the cone. This can be seen in Fig.~\ref{fig:companion_case}d representing the emitted radiation power $P_{\rm rad}$, which is a good tracer of particle acceleration since the mean Lorentz factor ($\langle\gamma\rangle$) is weighted by the local density. Particle acceleration continues to increase with radius until the end of the box, since magnetic reconnection continues to operate and radiative losses diminish with radius.

\begin{figure}[bthp]
\centering 
\includegraphics[width=\columnwidth, keepaspectratio]{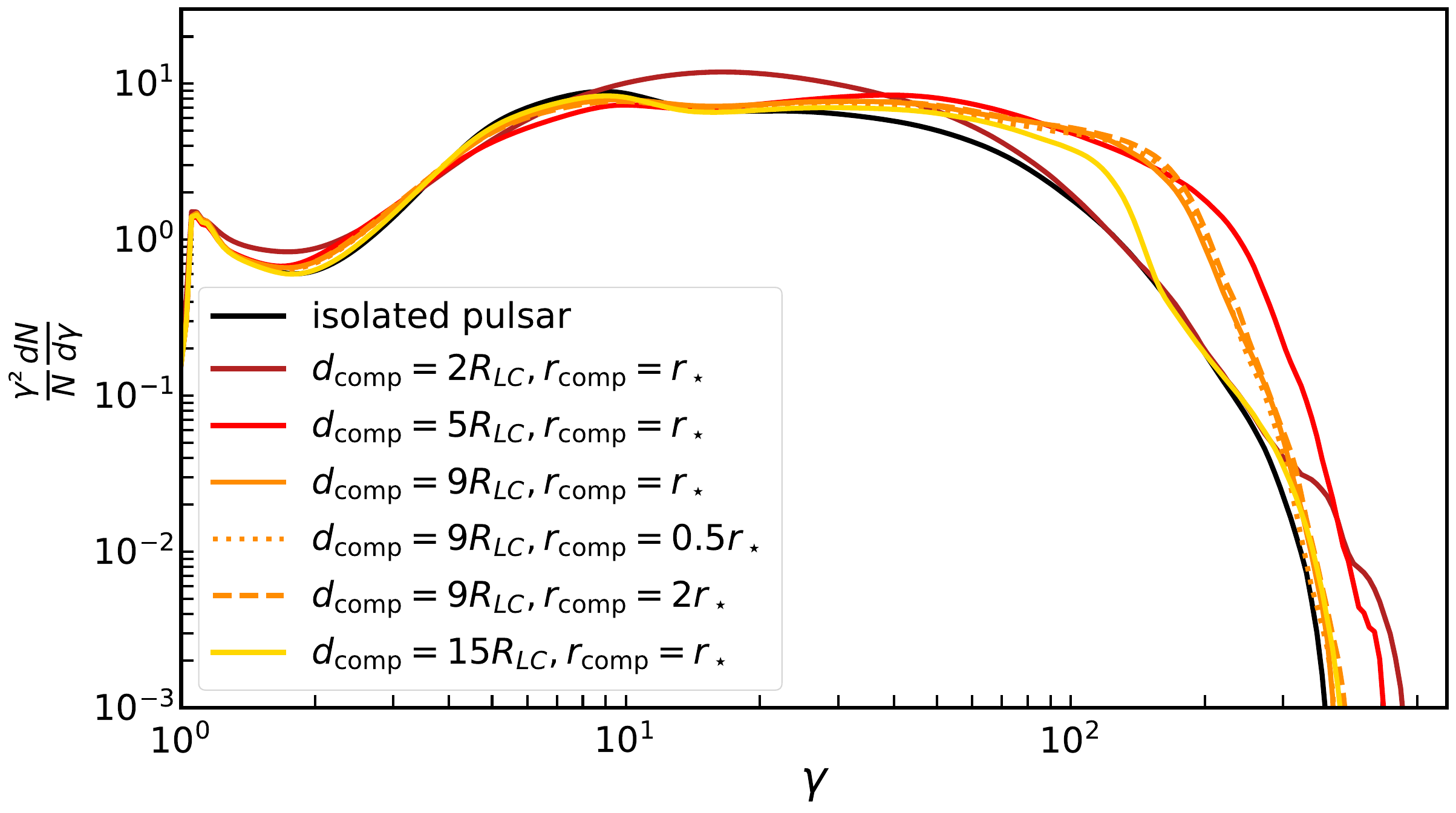}
\caption{Particles energy spectra for all runs, averaged over several spin periods. Spectrum of the isolated pulsar magnetosphere (black solid line) and spectra for all binary configurations (coloured solid lines). Orange dashed and dotted lines stand for different companion sizes.}\label{fig:spectra_particles}
\end{figure}

\begin{figure}[bthp]
\centering 
\includegraphics[width=\columnwidth, keepaspectratio]{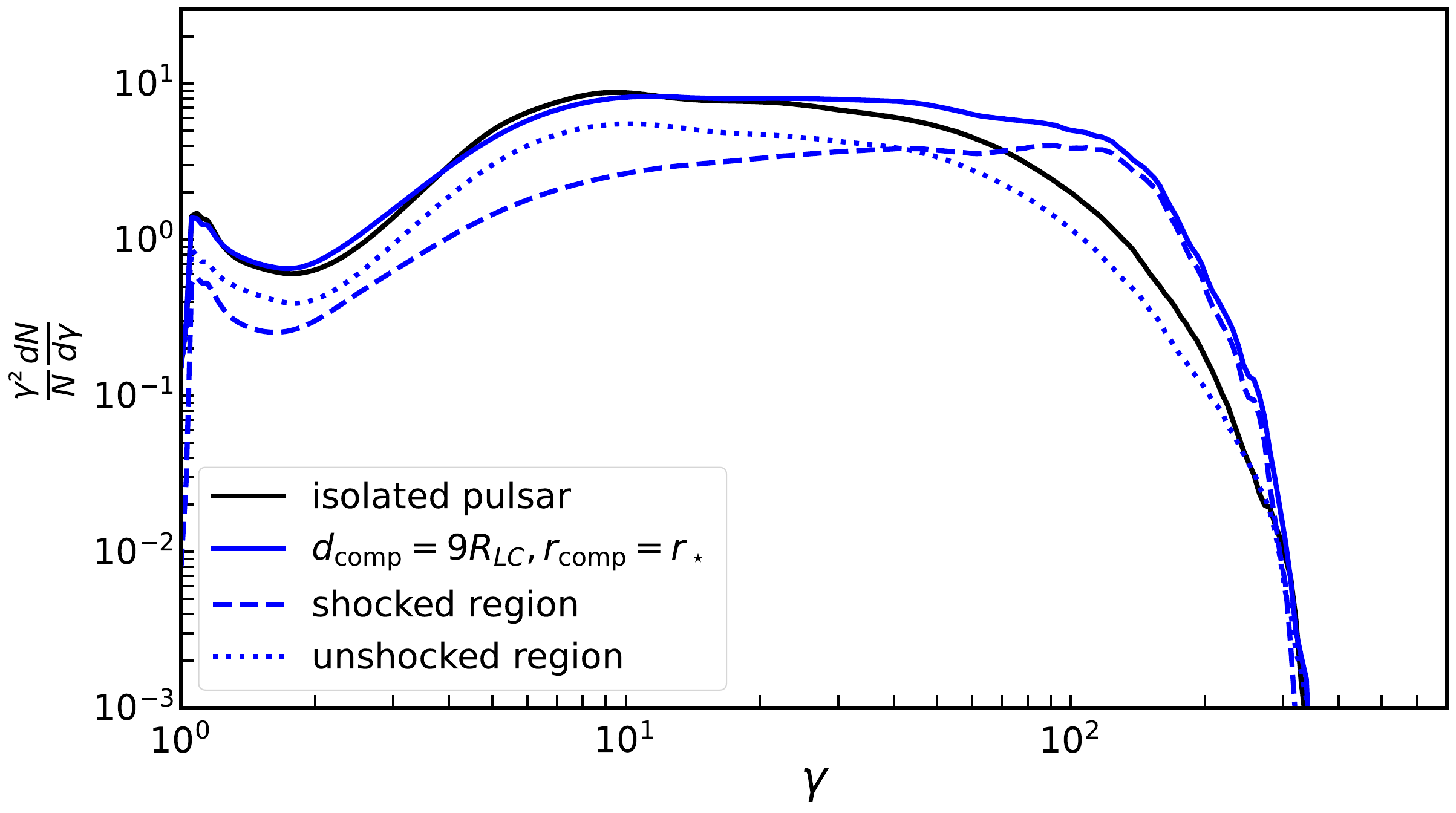}
\caption{Domain decomposition of the total particles energy spectrum (blue solid line) for the D9R1 run. Spectrum computed inside the shocked zone (blue dashed line) and spectrum computed outside of the shocked zone (blue dotted line). Comparison with the isolated pulsar case (black solid line). The spectrum alteration in the companion presence exclusively comes from the shocked zone.}\label{fig:spectra_particles_domains}
\end{figure}

Figure \ref{fig:spectra_particles} shows the particles energy spectra for all runs. The presence of the companion induces a localized bump in the spectrum, centered on different energies depending on the binary separation. While the separation of $2 R_{\rm LC}$ leads to a bump centered on $\gamma \sim 17$, the separation of $5 R_{\rm LC}$ seems to be the most favourable to reach the highest energies, with the bump being centered on $\gamma \sim 200$. The radius of the companion does not seem to play a significant role on the particles energy spectra (see orange lines on Fig.~\ref{fig:spectra_particles}). A domain decomposition based on the shocked zone (see Fig.~\ref{fig:spectra_particles_domains}) confirms that the observed hardening of the spectra entirely comes from the shocked part of the wind while, outside the shocked surface, the spectral shape is the same as for an isolated pulsar.

\subsection{Electromagnetic signature}\label{subsection:em_signature}

As previously mentioned (Section~\ref{subsection:radiation_modelling}), synchrotron losses account for most of the high-energy radiation in the studied system. As can be seen in Figure \ref{fig:ref_case} for the isolated pulsar configuration, synchrotron losses originate from the particles previously accelerated in the reconnection layers and confined inside the plasmoids along the current sheet. In particular, synchrotron emission is predominant in the inner parts of the magnetosphere where the magnetic field is stronger ($B \propto 1/r$) and the density higher ($n \propto 1/r^{2}$).

However, the presence of a companion adds a predominant contribution to the usual synchrotron losses. 
Indeed, when the current sheet hits the conductor surface, particles are further accelerated (see previous section) and the current sheet lights up again (see Fig. \ref{fig:current_sheet}). The current sheet is torn apart by the companion and the two separated branches resulting from this interaction flow radially beyond the companion, creating a hollow cone of light (as shown in Figure \ref{fig:companion_case}, panel d).  

\begin{figure}[bthp]
\centering 
\includegraphics[width=\columnwidth, keepaspectratio]{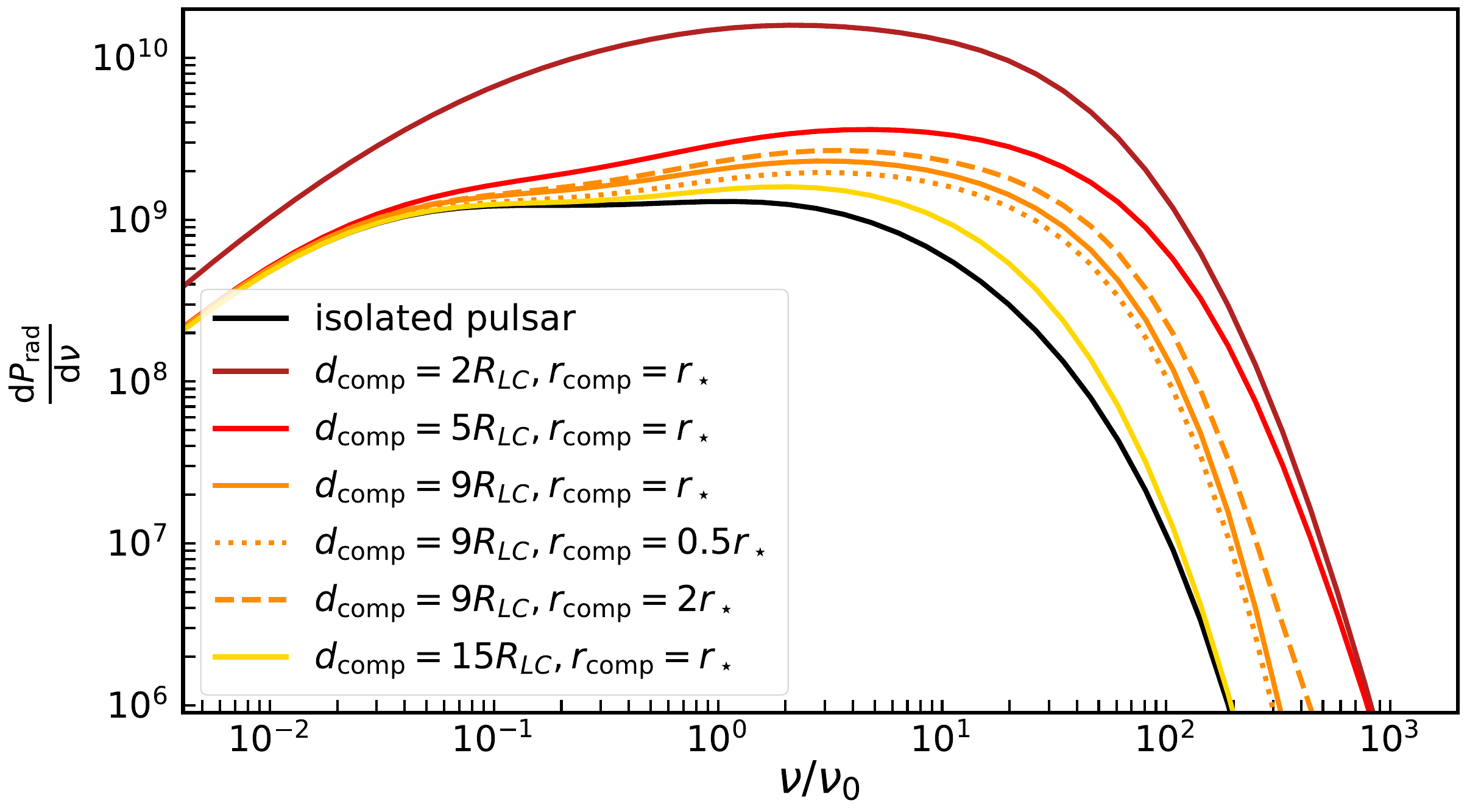}
\caption{Synchrotron spectra for all runs, averaged over several spin periods. Spectrum of the isolated pulsar magnetosphere (black solid line) and spectra for all binary configurations (coloured solid lines). Orange dashed and dotted lines stand for different companion sizes.}\label{fig:spectra_sync}
\end{figure}

\begin{figure}[bthp]
\centering 
\includegraphics[width=\columnwidth, keepaspectratio]{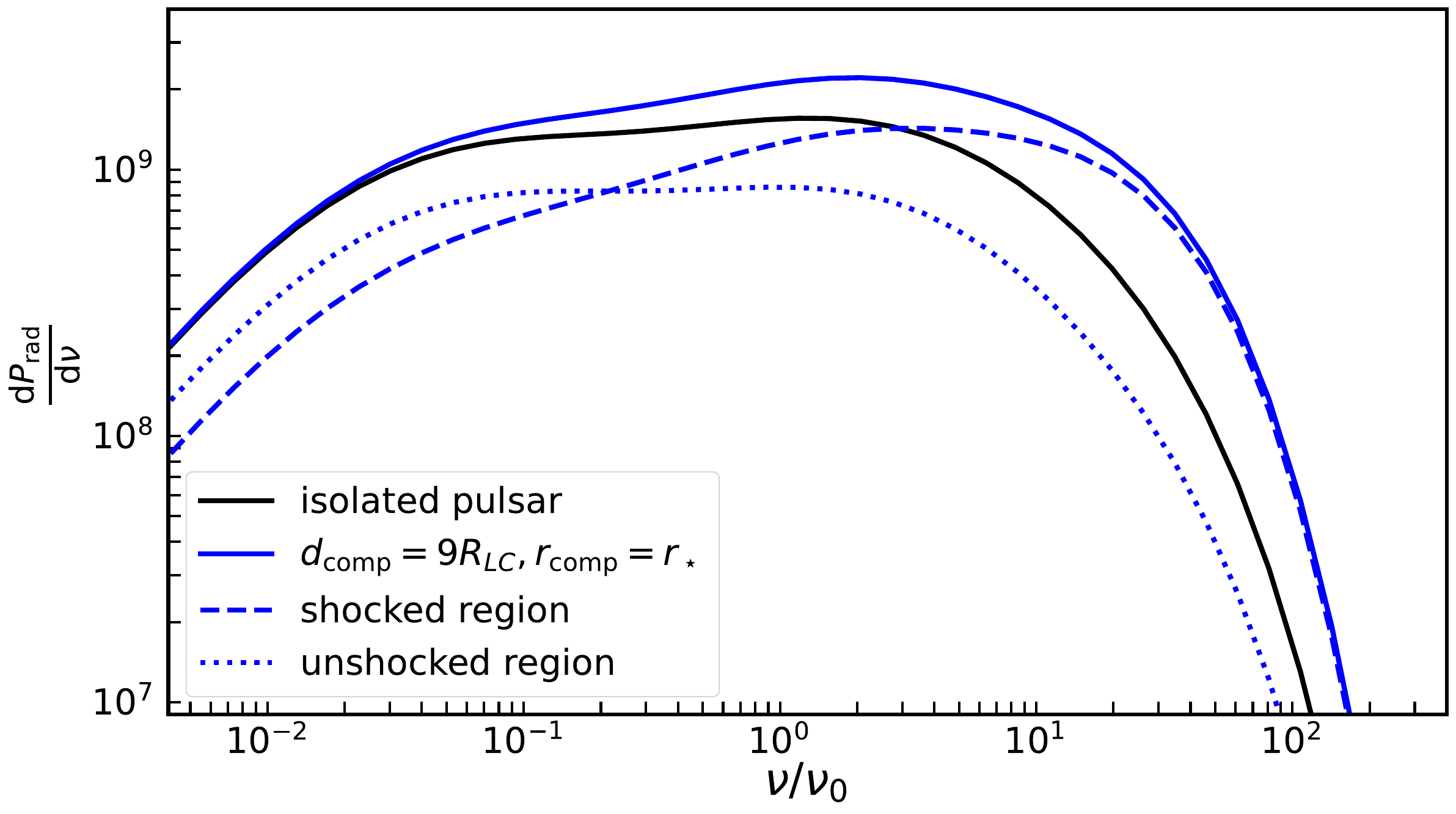}
\caption{Domain decomposition of the total synchrotron spectrum (blue solid line) for the D9R1 run. Spectrum computed inside the shocked zone (blue dashed line) and spectrum computed outside of the shocked zone (blue dotted line). Comparison with the isolated pulsar case (black solid line). The spectrum alteration in the companion presence exclusively comes from the shocked zone.}\label{fig:spectra_sync_domains}
\end{figure}

Figure \ref{fig:spectra_sync} compares the synchrotron spectra for all runs. As expected, for all binary separations, the presence of the companion significantly enhances synchrotron losses and shifts the peaks of the spectra to higher frequencies. Here, again, the separation $d_{\rm comp}=5 R_{\rm LC}$ seems to optimise the emission of synchrotron radiation peaking at a higher frequency compared to the other runs. The excess of synchrotron emission decreases with the binary separation. This mainly comes from the decrease of the density and the magnetic field strength with radius, despite the slower growth of $\gamma$ with radius. We notice a slight upward shift of the spectrum when doubling the companion size ($r_{\rm comp}=2 r_\star$), and a slighter downward shift when decreasing the companion radius ($r_{\rm comp}=0.5 r_\star$). Figure \ref{fig:spectra_sync_domains} makes a comparison, for the run D9R1, between the synchrotron spectrum emitted by the shocked part of the magnetosphere and the one emitted by the unshocked part. The unshocked part of the magnetosphere displays the same spectrum as the isolated pulsar magnetosphere, indicating that the additional contribution exclusively comes from the shocked region of the magnetosphere.

\begin{figure}[bthp]
\centering 
\includegraphics[width=\columnwidth, keepaspectratio]{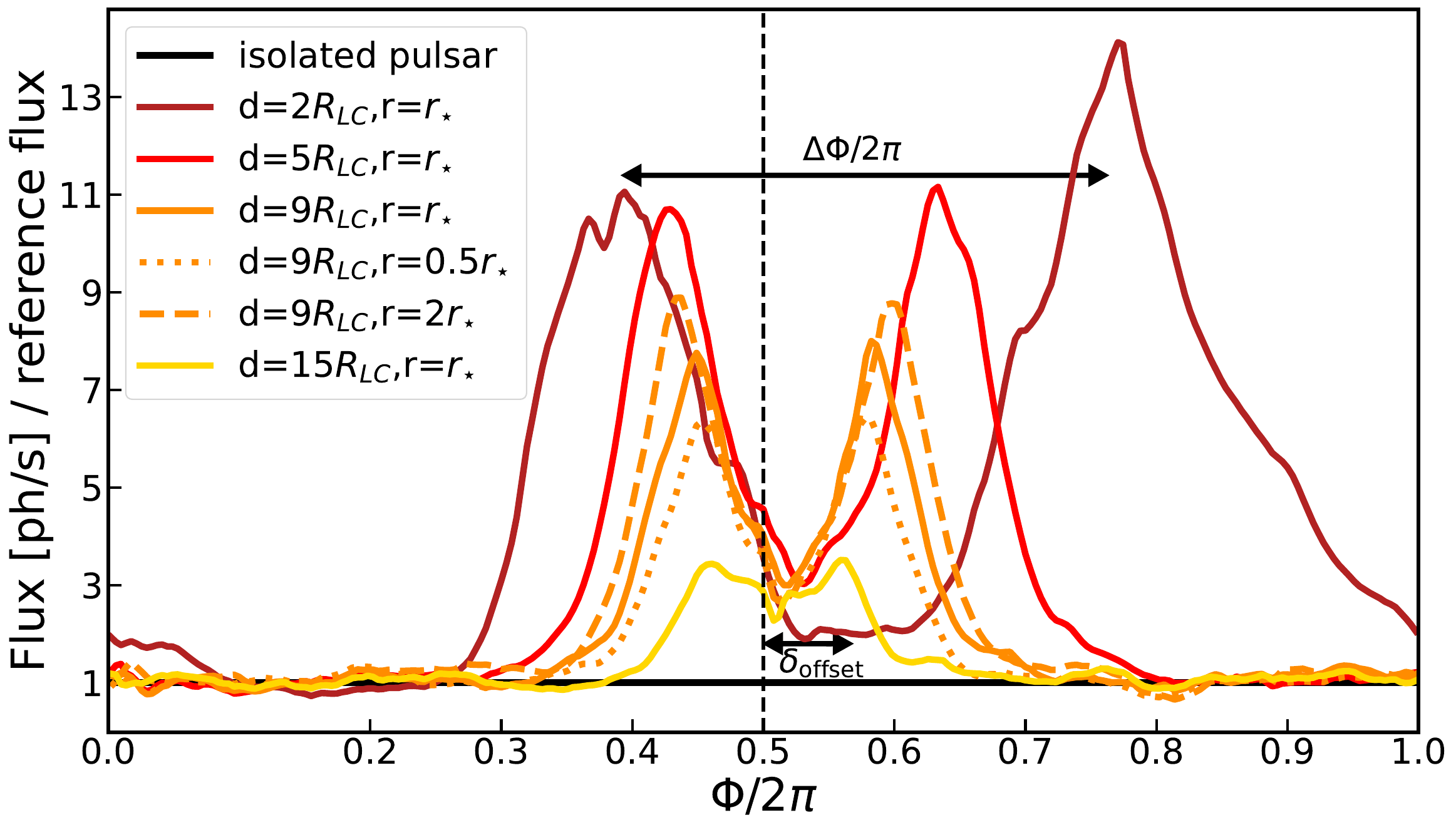}
\caption{High-energy ($\nu > \nu_0=3eB_\star/4\pi m c$) synchrotron light curves for all runs, as a function of the orbital phase, $\Phi$. Fluxes are integrated over the full polar angle range. Fluxes are normalized by the reference flux computed for the isolated pulsar magnetosphere (black solid line). Different colours stand for different binary separations. Orange dashed and dotted lines stand for different companion sizes. The pulsar-companion direction is given by the black dashed line. Two intense and broad peaks (corresponding to a hollow cone of light in 3D) are induced by the pulsar-companion interaction, which properties depend on the companion distance and size. The angle between peaks is referred to as $\Delta \Phi$ and the light cone tilt with respect to the pulsar-companion direction is designated by $\delta_{\rm offset}$.}\label{fig:light_curves}
\end{figure}

Orbital light curves for all runs are shown in Figure~\ref{fig:light_curves}. We represent the high-energy synchrotron emission taken above the fiducial synchrotron frequency $\nu_0=3 e B_\star/4\pi m c$, which was proven in previous studies to be an appropriate threshold for studying synchrotron emission from the current sheet (see \citealt{Cerutti_2016}). Light curves are integrated over the full polar angle range and are normalized by the reference flux, computed by averaging over $P_{\rm spin}$ the pulsed radiation emitted by the isolated pulsar magnetosphere (see Section~\ref{section:isolated_case}). All light curves present a significant enhancement of the radiation flux, with two broad peaks corresponding to the passing of the observer's line of sight through the edges of the hollow cone of emission (see Fig.~\ref{fig:companion_case}d). The peak height varies from $3.5$ to $14$ times the reference flux, depending on the companion separation and size. Higher companion sizes (orange lines on Fig. \ref{fig:light_curves}) induce higher peak intensities and higher orbital phases between peaks. The binary separation has an impact on the intensity of the peaks, their widths, the separation between them ($\Delta \Phi$), and the light cone orientation with respect to the pulsar-companion direction ($\delta_{\rm offset}$).

\begin{figure}[bthp]
\centering 
\includegraphics[width=\columnwidth, keepaspectratio]{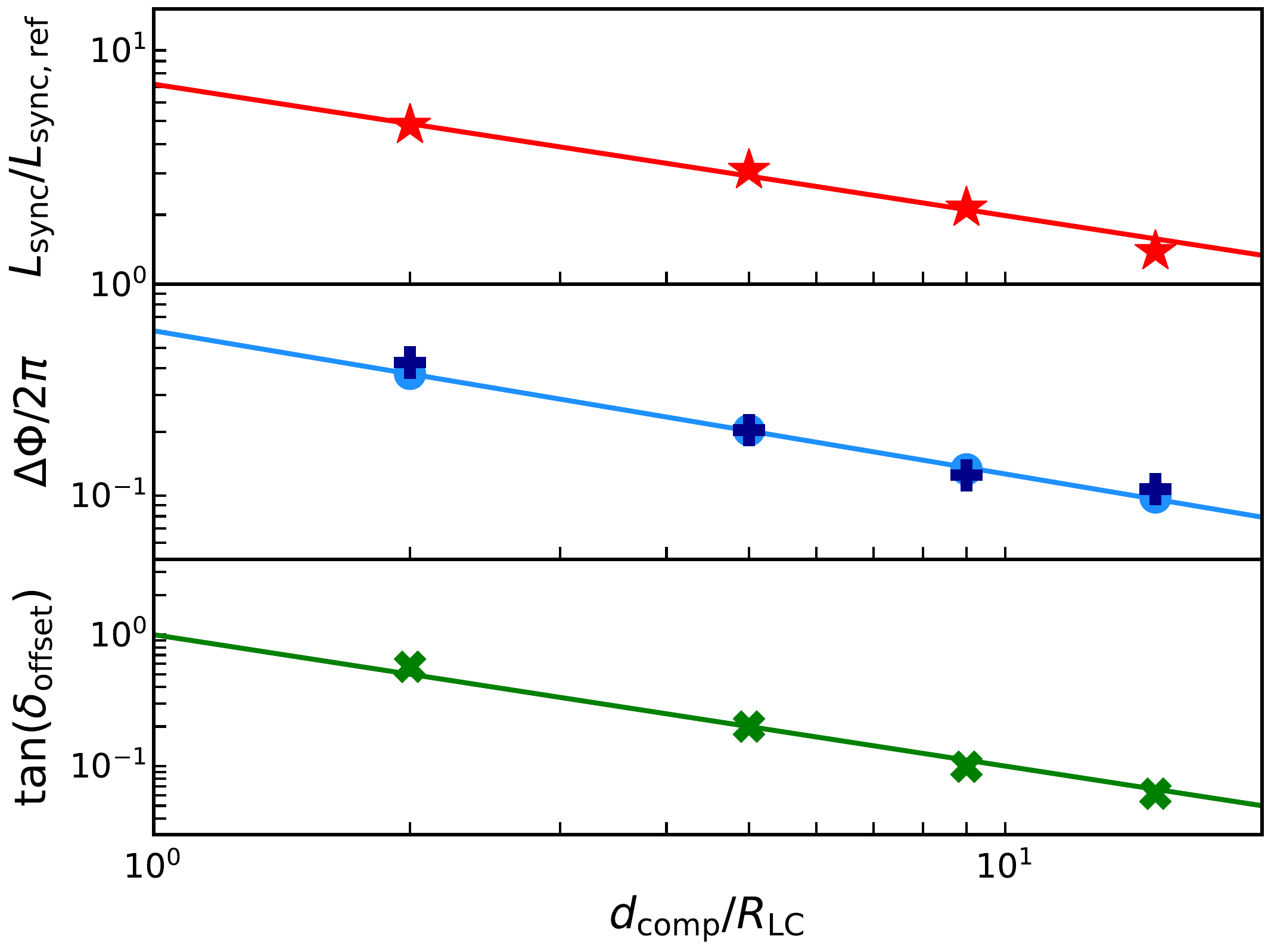}
\caption{Light curves properties as a function of the pulsar-companion separation. \textit{Upper panel:} integrated synchrotron power over the whole orbit (red points), normalized by the power computed for the isolated pulsar. The total synchrotron power scales as $d_{\rm comp}^{-0.56}$ (red line). \textit {Middle panel:}  the orbital phase difference between peaks (light blue points) scales as $d_{\rm comp}^{-0.68}$ (light blue line). We plot the corresponding $1/\Gamma$ factors normalized by a factor $0.55$, computed before the companion for each run (dark blue crosses). \textit{Lower panel:} Angular offsets of the emitting cone with respect to the pulsar-companion direction (green points) and ratio of the wind drift velocity components $V_\phi/V_r=R_{\rm LC}/r$ (green line) given by the monopole analytical solution \citep{1973ApJ...180L.133M}.} \label{fig:light_curves_dependences}
\end{figure}

The exact dependency between these parameters and the pulsar-companion separation is presented in Figure \ref{fig:light_curves_dependences}. The upper panel shows that the total synchrotron power emitted over an orbit scales as $d_{\rm comp}^{-0.56}$ (red line). The orbital phases between both peaks (light blue points on the middle panel of Fig.~\ref{fig:light_curves_dependences}) decrease with radius as $d_{\rm comp}^{-0.68}$ (light blue line). We note that they scale similarly to the corresponding $1/\Gamma$ factors, computed for each run at the apex of the shocked cone (dark blue points on the middle panel of Fig.~\ref{fig:light_curves_dependences}), suggesting that the peak separation is shaped by relativistic beaming effects. In Figure~\ref{fig:light_curves_dependences}, the points corresponding to the $1/\Gamma$ scaling are normalized by a factor of $0.55$. This discrepancy can be explained by the values of $\Gamma$ inferred from the averaged maps (shown in Fig.~\ref{fig:gamma_bulk}), which erase any inhomogeneities. Indeed, as can be seen in Figure~\ref{fig:companion_case}c, $\Gamma$ can locally be amplified by up to a factor of $2$ inside the current sheet near the companion. The lower panel demonstrates that the angular offsets of the light hollow cones with respect to the pulsar-companion direction are determined by the plasma drift velocity at the companion radius. Indeed, the values of ${\rm tan} (\delta_{\rm offset})$ for each companion separation (green crosses) are given by the corresponding drift velocity ratios $V_\phi/V_r=R_{\rm LC}/r$ (green line) taken at the companion location and computed according to the monopole analytical solution (derived in~\citealt{1973ApJ...180L.133M}). We note that the shocked cones represented in Figure~\ref{fig:gamma_bulk} are tilted, for each run, by exactly the corresponding angle $\delta_{\rm offset}$.

\section{Discussion and conclusion}\label{section:discussion}

In this work, we study the interaction of a pulsar wind with a similarly sized companion to assess the extent to which the pulsar magnetosphere reshapes itself. We probe the enhancement and location of particle acceleration to predict the non-thermal radiation originating from such systems. We show that the interaction between a pulsar wind and a companion alters significantly the dynamical and energetic properties of the wind. The pulsar wind is slowed down rather isotropically if the companion lies within the fast magnetosonic point. Otherwise, when the companion is settled beyond the fast magnetosonic point, a shock is established and all the perturbations are advected in a cone behind the companion. Each time the outflowing wind current sheet impacts the companion surface, we observe a forced reconnection that leads to a significant enhancement of particle acceleration. Reaccelerated particles form a hollow cone behind the companion. By doing so, they induce an orbital-modulated hollow cone of high-energy synchrotron radiation, whose intensity and aperture depend on the orbital separation and the size of the companion. Hence, non-thermal radiation from such systems is significantly enhanced as well, compared to the reference pulsed flux of an isolated pulsar magnetosphere.

The 2D framework used here may introduce some artifacts. First, the magnetic field is confined in the equatorial plane, which may have an impact on the rearrangement of the field lines. It is worth noting that our results give an upper limit in terms of particle acceleration and high-energy emission. Indeed, the equatorial configuration necessarily implies that all the magnetic field lines accumulating ahead of the companion reconnect, possibly overestimating the amount of dissipation into kinetic energy. In addition, in 2D, particles accelerated by magnetic reconnection escape the X-points and are all trapped by the neighbouring plasmoids, however, a third dimension would allow for the leakage of the accelerated particles outside of plasmoids \citep{2021ApJ...922..261Z, 2023ApJ...959..122C}. This could slightly alter the shape of the light hollow cone. As mentioned at the beginning of this study, the scale separation that we have been able to reach in these PIC simulations is reduced by several orders of magnitude compared to a realistic pulsar magnetosphere. This numerical limitation constrains the frequency range of the high-energy radiation, leaving many uncertainties on the spectral limits and the spectral shape -- but not on the total energy flux that is conserved. The presence of a companion wind would imply a bigger effective surface, but should not radically alter the nature of the shock. A slight asymmetry between the two peaks appears in the light curves. Taking into account the orbital motion of the companion could potentially exacerbate this asymmetry. 

We predict an orbital-modulated emission that could originate from the orbital motion of asteroids, planets, or neutron stars around the pulsar, with a higher chance of observing such systems if seen close to edge-on. The characteristic frequency of the non-thermal radiation is expected to fall in the soft gamma-ray band. We therefore expect these transients to be observable only on galactic distances. While not investigated in this work, radio counterparts are expected in addition to the high-energy non-thermal radiation. Indeed, in the wake of the companion where the wind is altered, we observe the propagation of small-scale fast magnetosonic modes. According to \citet{Lyubarsky_2019} and \citet{Philippov_2019}, the collision of plasmoids with each other and with the magnetic field perturbates the magnetic field and induces short fast magnetosonic pulses that eventually escape the plasma as radio waves. Other coherent mechanisms of radio emission could also be at play, such as the cyclotron maser instabilility that was initially retained to describe fast radio bursts due to an outflowing pulsar wind crossing Alfvén wings \citep{Neubauer_1980,Mottez_2011b} formed behind a companion \citep{Heyvaerts_2012,Mottez_2012,Mottez_2014,Mottez_2020,Decoene_2021}. Galactic radio counterparts would be of significant interest, especially in the light of the recently discovered galactic long-period radio transients \citep{Walker_2022,Walker_2023}. These periodic radio transients are characterized by very long periods, ranging from $10$s to $1000$s. Their sources have not been uniquely identified yet; however, binary systems including a pulsar have been noted as possible source candidates \citep{Rea_2022,Rea_2024}.

\begin{acknowledgements}
This project has received funding from the European Research Council (ERC) under the European Union’s Horizon 2020 research and innovation program (Grant Agreement No. 863412). Computing resources were provided by TGCC under the allocation A0150407669 made by GENCI.
\end{acknowledgements}

\bibliographystyle{aa}
\bibliography{main}

\end{document}